\documentclass[seceq]{ptptex}
\usepackage{wrapft}
\usepackage{amsmath}
\usepackage{amssymb}
\usepackage{bm}
\notypesetlogo                       

\title{
Residual Entropy of the Mott Insulator with No Symmetry Broken
}
\author{
Fusayoshi J. \scshape{Ohkawa}\footnote{E-mail: fohkawa@phys.sci.hokudai.ac.jp} 
}
\inst{
Division of Physics, Hokkaido University, Sapporo 060-0810, Japan
}

\recdate{
March 14, 2012; revised May 7, 2012}
\abst{        
The half-filled ground state of the Hubbard model on the hypercubic lattice in $D$ dimensions is studied  
by the Kondo-lattice theory, which is none other than the $1/D$ expansion theory, but within the constrained Hilbert subspace where no symmetry is allowed to be broken.
A gap can open in the single-particle excitation spectrum if and only if the residual entropy or entropy at $T=+0\hskip2pt$K is nonzero.
The Mott insulator with no symmetry broken, if it is possible, is characterized by nonzero residual entropy or nonzero entropy at $T=+0\hskip2pt$K.  
This conclusion is consistent with Brinkman and Rice's theory and the dynamical mean-field theory.  According to the well-known argument based on the Bethe-ansatz solution, on the other hand, the half-filled ground state in one dimension is the Mott insulator although its residual entropy per unit cell is vanishing in the thermodynamic limit.
Two possible explanations are given for the contradiction between the present paper and the well-known argument.
}
\PTPindex{345, 346, 397}   

\begin{document}
\maketitle 

\section{Introduction}
\label{SecIntroduction}
Mott localization or the Mott insulator is a longstanding crucial issue \cite{mott}.
Whether magnetism is itinerant-electron or local-moment magnetism is directly related to the physics of the Mott insulator.
High-temperature (high-$T_c$) superconductivity occurs in the vicinity of the phase of a local-moment type of antiferromagnetic  insulator \cite{bednortz}.
Thus, it is anticipated that the mechanism of high-$T_c$ superconductivity is closely related to the physics of the Mott insulator \cite{Anderson-SC,FJO-SC1,FJO-SC2}.

For example, we consider the Hubbard model at the absolute zero Kelvin.
According to Hubbard's theory,\cite{Hubbard1,Hubbard2} 
when the on-site repulsion $U$ is as large as or larger than the bandwidth $W$,
the band splits into the upper and lower Hubbard bands, i.e., the Hubbard gap opens. 
According to Gutzwiller's theory,\cite{gutzwiller1,gutzwiller2,gutzwiller3} 
when $U\gtrsim W$ and electrons are almost half-filled,
electrons are renormalized into heavy quasi-particles.
According to a slave-boson theory,\cite{midband} 
a midband appears at the chemical potential between the upper and lower Hubbard bands; the midband is none other than a band of heavy quasi-particles in Gutzwiller's theory, so that it is called the Gutzwiller band.
According to Brinkman and Rice's theory,\cite{brinkman} which is based on Gutzwiller's theory, 
if electrons are exactly half-filled, a metal-insulator transition occurs at a critical $U_c\simeq W$. In an insulating phase at $U> U_{c}$, the Gutzwiller band disappears and the Hubbard gap becomes a complete gap.
According to these theories, the ground state is a metal for any $U$ unless electrons are exactly half-filled and, therefore, 
the ground state can be the Mott insulator if and only if electrons are exactly half-filled and the onsite $U$ is large enough.

%

According to Brinkman and Rice's theory,\cite{brinkman} the specific-heat coefficient diverges as the onsite $U$ approaches the critical $U_c$ from the metallic or nongapped phase; and no symmetry is broken in the insulating or gapped phase. 
A combination of these two properties means or at least implies that the residual entropy is nonzero in the gapped phase, although it is unlikely that the third law of thermodynamics is broken. 
Whether the residual entropy is zero or nonzero in the gapped phase is a crucial issue to be elucidated.

Either Hubbard's or Gutzwiller's theory is under single-site approximation (SSA).
If all the single-site terms are rigorously considered by a theory, the theory is the supreme SSA (S$^3$A) theory. 
In the limit of $D\rightarrow +\infty$, where $D$ is the spatial dimensionality, multisite terms vanish  except for certain types of conventional Weiss mean field,%
\footnote{
In a single-band model such as the Hubbard model, three types of conventional Weiss mean field are of the zeroth order in $1/D$: spin density wave (SDW) or magnetism, charge density wave (CDW), and Bardeen-Cooper-Schrieffer (BCS) or isotopic $s$-wave superconductivity.
In a multiband model, an orbital order in the charge or spin channel is also of the zeroth order in $1/D$.}
which is a multisite effect.
Thus, the S$^3$A is rigorous in the limit of $D\rightarrow +\infty$,\cite{Metzner,Muller-H1,Muller-H2,Janis}
but within the constrained Hilbert subspace where no symmetry is allowed to be broken.
A problem of rigorously calculating all the single-site terms under S$^3$A can be substituted for a self-consistent Kondo problem,\cite{Mapping-1,Mapping-2,Mapping-3,georges}
in which the Anderson model to be solved has to be self-consistently determined with the single-site terms. A dynamical mean-field theory\cite{georges,RevModDMFT} (DMFT)
or a dynamical coherent-potential approximation (DCPA) theory\cite{dcpa} is none other than the S$^3$A theory.


Under DMFT or S$^3$A,
a metal-insulator transition occurs at $U$ as large as Brinkman and Rice's $U_c$.\cite{kotliar,moeller, bulla} 
Since hysteresis appears, it is a discontinuous transition between the metallic and insulating phases at $T=0\hskip2pt$K.
Since no jump occurs in the internal energy of Brinkman and Rice's theory, a plausible explanation for the hysteresis is that the residual entropy jumps at the transition.
It is interesting to elucidate which is discontinuous at the transition point and is responsible for the hysteresis, the residual entropy, the internal energy, or some thing else.

Even if multisite terms are approximately or rigorously considered beyond S$^3$A,
a problem of rigorously calculating all the single-site terms can also be substituted for another self-consistent Kondo problem,\cite{Mapping-1,Mapping-2,Mapping-3,toyama} in which the Anderson model to be solved has to be self-consistently determined with multisite terms as well as the single-site terms.
Since the Anderson model is an effective Hamiltonian for the Kondo problem, a theory for treating either the self-consistent Kondo problem
under or beyond S$^3$A is none other than a Kondo-lattice theory;
\cite{Mapping-1,Mapping-2,Mapping-3,toyama} 
if multisite terms are considered beyond S$^3$A,  
the Kondo-lattice theory is none other than a $1/D$ expansion theory.
In the present paper, we list all the possible types of gapped phase of the Hubbard model under and beyond S$^3$A and then we examine on the basis of the Kondo-lattice or $1/D$ expansion theory whether the residual entropy is zero or nonzero in each of them; 
we consider only possible phases within the constrained Hilbert subspace in which no symmetry is allowed to be broken, so that
the possibility of the Mott insulator with symmetry broken or one with a hidden order is out of scope.

The present paper is organized as follows:
Preliminaries are given in \S\ref{SecPreliminary}.
In \S\ref{SecDMFT}, nongapped and gapped phases under S$^3$A are studied by the Kondo-lattice theory in order to confirm that, under S$^3$A, the Kondo-lattice theory is equivalent to DMFT.
In \S\ref{SecKLT}, only possible complete-gapped phases beyond S$^3$A  are studied by the Kondo-lattice theory.
Discussion is given in \S\ref{SecDiscussion}.
Conclusion is given in \S\ref{SecConclusion}.
Anomalies of possible gapped phases are studied in Appendix\hskip2pt\ref{SecAnomaly}.
The entropy at $T=+0\hskip2pt$K of possible complete-gapped phases is studied 
in Appendix\hskip2pt\ref{SecResidualEntropy}.
Possible gapped phases in a nonlattice or continuous model, such as the Tomonaga-Luttinger model,\cite{Tomonaga,Luttinger,Mattis} are studied in Appendix\hskip2pt\ref{SecContinousModel}.

\section{Preliminaries}
\label{SecPreliminary}
\subsection{Model}
\label{SecModel}
We consider the Hubbard model on a hypercubic lattice in $D$ dimensions:
\begin{equation}\label{EqHubbardH}
{\cal H} = 
- \sum_{ij\sigma} 
t_{ij}  d_{i\sigma}^\dag d_{j\sigma}
+U \sum_i n_{i\uparrow}n_{i\downarrow},
\end{equation}
where 
$n_{i\sigma}=d_{i\sigma}^\dag d_{i\sigma}$ and the other notations are conventional. 
The periodic boundary condition is assumed.
When $U=0$, the dispersion relation of an electron is given by
\begin{align}
E({\bm k}) &=
- \frac1{L}\sum_{ij}t_{ij} e^{i{\bm k}\cdot({\bm x}_i-{\bm x}_j)}, 
\end{align} 
where ${\bm k}=(k_1, k_2, \cdots, k_D)$ is the wave number of the electron, ${\bm x}_i =(i_1, i_2, \cdots, i_D)a$, with $a$ being the lattice constant, is the position of the $i$th unit cell, and $L$ is the number of unit cells.
The hypercubic lattice is decomposed into two sublattices: one composed of unit cells with $\sum_{\nu=1}^{D}i_\nu$ being even and the other composed of unit cells with $\sum_{\nu=1}^{D}i_\nu$ being odd. For simplicity, we assume that the transfer integral $t_{ij}$ is nonzero only between different sublattices, i.e., the Hubbard model is on a bipartite lattice. 
The density of states per unit cell is defined  by
\begin{align}
\rho_0(\varepsilon) &=
\frac1{L}\sum_{\bm k} \delta\bigl[\varepsilon- E({\bm k})\bigr],
\end{align}
which is called the bare density of states in this paper. 
The band center of $E({\bm k})$ or $\rho_0(\varepsilon)$ is denoted by $\epsilon_d$; $\epsilon_d=0$ for the model of this paper.
A hypersurface in the wave-number space, ${\bm k}$ on which is denoted as ${\bm k}_{\rm F}$, is defined by
%
\begin{align}\label{EqHyper-surface}
E({\bm k}_{\rm F})=\epsilon_d=0.
\end{align}
%
Since $E({\bm k})$ is symmetrical with respect to the hypersurface, $\rho_0(\varepsilon)$
is symmetrical with respect to $\epsilon_d=0$:
$\rho_0(\varepsilon)= \rho_0(-\varepsilon)$.
We assume that the chemical potential $\mu$ is exactly given by $\mu = U/2$ and that the Hubbard model is exactly half-filled. 
In general, the particle-hole symmetry exists in a half-filled model on a bipartite lattice.
The particle-hole symmetry is crucial in this study.

The transfer integral $t_{ij}$ has to include the dimensional factor $D$ in order for the effective bandwidth of $E({\bm k})$ to be finite for any $D$.\cite{Metzner,Muller-H1,Muller-H2,Janis}
For example, if $t_{ij}$ is nonzero only between nearest neighbors, $t_{ij}=t/\sqrt{D}$ between them. It follows that
\begin{align}
E({\bm k}) &
=-2 \frac{t}{\sqrt{D}} \sum_{\nu=1}^{D} \cos (k_\nu a). 
\end{align} 
The absolute bandwidth of $E({\bm k})$ is given by $W=4|t|\sqrt{D}$. 
Since $(1/L)\sum_{\bm k} E({\bm k})=0$ and  
$(1/L)\sum_{\bm k} E^2({\bm k})= 2t^2$, the effective bandwidth of $ E({\bm k})$ is $O\left(|t|\right)$ for any $D$.
If $t_{ij}$ is of finite range, either the absolute or effective bandwidth is finite for any finite $D$.
We call any model of this type for a finite $D$ a finite bandwidth model.


If $t_{ij}$ is of infinitely long range, it is possible that $E({\bm k})$ is given by
\begin{align}
E({\bm k}) = 
\sum_{\nu=1}^{D} \frac{|t_\infty|}{D} \eta(k_\nu a),
\end{align}
where
\begin{align}
\eta(x) = \left\{\begin{array}{cc}
\displaystyle 
\phantom{-}\tan\left(x - \pi/2\right), & \displaystyle 0< x < \pi, \vspace{3pt} \\
\displaystyle 
-\tan\left(x + \pi/2\right), & \displaystyle -\pi< x < 0 .\\
\end{array} \right.  
\end{align}
The bare density of states is of a Lorentzian type for any $D$:  
\begin{align}\label{EqDOS-Lorentzian} 
\rho_0(\varepsilon)
= \frac1{L}\sum_{\bm k} \delta\bigl[\varepsilon-E({\bm k})\bigr]
= \frac1{\pi} \frac{|t_\infty|}{\varepsilon^2+ |t_\infty|^2}.
\end{align}
%
We call this model an infinite bandwidth model.

\subsection{Kondo-lattice theory}
\label{SecKL-theory}
We constrain the Hilbert space within the subspace where no symmetry is allowed to be broken.
We first assume that $L$ is large but still finite and that the temperature is nonzero such that $T>0\hskip2pt$K; the thermodynamic limit of $L\rightarrow +\infty$, if necessary, followed by the low-temperature limit of $T\rightarrow 0\hskip2pt$K is taken last.

The site representation is convenient for formulating the Kondo-lattice theory. 
The Green function and self-energy in the grand canonical ensemble are denoted by $R_{ij;\sigma}({i}\varepsilon_l)$ and $\Sigma_{ij;\sigma}({i}\varepsilon_l)$, respectively,
where $\varepsilon_l=(2l+1)\pi k_{\rm B}T$, with $l$ being an integer, is a fermionic energy. We consider a skeleton Feynman diagram for $\Sigma_{ij;\sigma}({i}\varepsilon_l)$, but with each electron line replaced by a renormalized line of $R_{ij;\sigma}({i}\varepsilon_l)$.
If only lines of $U$ and site-diagonal $R_{ii\sigma}({i}\varepsilon_l)$ appear in it, it is a single-site diagram;
if at least one line of site-off-diagonal $R_{i\ne j;\sigma}({i}\varepsilon_l)$ appears in it, it is a multisite diagram.
All the Feynman diagrams can be classified into the single-site and multisite ones without any ambiguity. Thus,
%
\begin{align}
\Sigma_{ij;\sigma}({i}\varepsilon_l) =
\Sigma_{i;\sigma}({i}\varepsilon_l)\delta_{ij}
+ \Delta\Sigma_{ij:\sigma}({i}\varepsilon_l),
\end{align}
where $\Sigma_{i;\sigma}({i}\varepsilon_l)$ is the single-site self-energy, $\delta_{ij}$ is the Kronecker delta, and $\Delta\Sigma_{ij:\sigma}({i}\varepsilon_l)$ is the multisite self-energy.
Since the site-diagonal $R_{ii;\sigma}({i}\varepsilon_l)$ and the single-site $\Sigma_{i;\sigma}({i}\varepsilon_l)$ do not depend on the site index $i$, they are simply denoted by $R_{\sigma}({i}\varepsilon_l)$ and $\Sigma_{\sigma}({i}\varepsilon_l)$, respectively, in the following part of this paper:


The single-site $\Sigma_{\sigma}({i}\varepsilon_l)$ is of the zeroth or leading order in $1/D$, while the multisite $\Delta\Sigma_{ij:\sigma}({i}\varepsilon_l)$ is of higher order in $1/D$.\cite{Metzner,Muller-H1,Muller-H2,Janis}
Any type of conventional Weiss mean field is a multisite effect.
If no multisite term is considered, the Kondo-lattice theory is reduced to S$^3$A, which is also DMFT or DCPA.
In this paper, the multisite $\Delta\Sigma_{ij:\sigma}({i}\varepsilon_l)$ is considered, but no conventional Weiss mean field is considered.

If the multisite $\Delta\Sigma_{ij:\sigma}({i}\varepsilon_l)$ is ignored under S$^3$A or if it is approximately or rigorously considered beyond S$^3$A, the single-site $\Sigma_\sigma({i}\varepsilon_l)$ is exactly equal to the local self-energy $\tilde{\Sigma}_\sigma({i}\varepsilon_l)$ of the Anderson model that is determined by solving the self-consistent Kondo problem, as discussed below. The mapping of $\Sigma_\sigma({i}\varepsilon_l)$ to $\tilde{\Sigma}_\sigma({i}\varepsilon_l)$, such as
\begin{align}\label{EqMapS}
\Sigma_\sigma({i}\varepsilon_l) &=
\tilde{\Sigma}_\sigma({i}\varepsilon_l),
\end{align}
is the key of the Kondo-lattice theory. 
The Anderson model is defined by
\begin{align}\label{EqAnderson}
\tilde{\cal H} &=
\tilde{\epsilon}_d \sum_{\sigma}\tilde{n}_{d\sigma}
+ \sum_{{\bm k}\sigma} \tilde{E}_c({\bm k}) \tilde{c}_{{\bm k}\sigma}^\dag \tilde{c}_{{\bm k}\sigma}
%
+ \frac1{\sqrt{ \tilde{L} }} \sum_{{\bm k}\sigma} \left(
\tilde{V}_{\bm k}\tilde{c}_{{\bm k}\sigma}^\dag \tilde{d}_\sigma
+ \tilde{V}_{\bm k}^* \tilde{d}_\sigma^\dag \tilde{c}_{{\bm k}\sigma}\right) 
+ \tilde{U} \tilde{n}_{d\uparrow} \tilde{n}_{d\downarrow},
\end{align} 
where 
$\tilde{n}_{d\sigma}=\tilde{d}_{\sigma}^\dag \tilde{d}_{\sigma}$ and the notations are conventional except for the use of tildes.
The local Green function for localized electrons in the Anderson model is given by
\begin{align}\label{EqGreenAM}
\tilde{G}_{\sigma}({i}\varepsilon_l) &=
\frac1{\displaystyle
{i}\varepsilon_l + \tilde{\mu} \hskip-1pt - \hskip-1pt \tilde{\epsilon}_d - \tilde{\Sigma}_\sigma({i}\varepsilon_l) - \tilde{\Gamma}({i}\varepsilon_l) }, 
\end{align}
where $\tilde{\mu}$ is the chemical potential, $\tilde{\Sigma}_\sigma({i}\varepsilon_l)$ is the local self-energy, and
\begin{subequations}\label{EqHybridGamma0}
\begin{align}\label{EqHybridGamma}
\tilde{\Gamma}({i}\varepsilon_l) &
= \frac1{\pi} \int_{-\infty}^{+\infty}\hskip-8pt d \epsilon \hskip1pt
\frac{\tilde{\Delta}(\epsilon)}{{i}\varepsilon_l - \epsilon },
\end{align}
with
\begin{equation}\label{EqDelta}
\tilde{\Delta}(\varepsilon) =
\frac{\pi}{\tilde{L}} \sum_{\bm k} |\tilde{V}_{\bm k}|^2 
\delta\bigl[\varepsilon+\tilde{\mu}- \tilde{E}_c({\bm k})\bigr] ,
\end{equation}
\end{subequations}
is the hybridization term with conduction electrons; and $\tilde{\Delta}(\varepsilon) =- {\rm Im} \tilde{\Gamma}(\varepsilon+{i}0)$.
Although the Anderson model is characterized by $\tilde{\mu}-\tilde{\epsilon}_d$, $\tilde{U}$, $\tilde{E}_c({\bm k})$, and $\tilde{V}_{\bm k}$, 
it is essentially uniquely characterized only by $\tilde{\mu}-\tilde{\epsilon}_d$, $\tilde{U}$, and $\tilde{\Delta}(\varepsilon)$. In the Kondo-lattice theory, not only the local $\tilde{\Sigma}_\sigma({i}\varepsilon_l)$  of the Anderson model and the multisite $\Delta\Sigma_{ij:\sigma}({i}\varepsilon_l)$ of the Hubbard model, but also the $\tilde{\mu}-\tilde{\epsilon}_d$, $\tilde{U}$, and $\tilde{\Delta}(\varepsilon)$ of the Anderson model have to be self-consistently calculated or determined to satisfy Eq.\hskip2pt(\ref{EqMapS}); the Anderson model determined in this way is called the mapped Anderson model.


Even if any arbitrary diagram for $\Sigma_\sigma({i}\varepsilon_l)$ of the Hubbard model is considered, the very same one for $\tilde{\Sigma}_\sigma({i}\varepsilon_l)$ of the Anderson model is obtained.
This one-to-one correspondence between the diagrams for $\Sigma_\sigma({i}\varepsilon_l)$ and $\tilde{\Sigma}_\sigma({i}\varepsilon_l)$ is unique, and there is no lonely diagram in either $\Sigma_\sigma({i}\varepsilon_l)$ or $\tilde{\Sigma}_\sigma({i}\varepsilon_l)$.
In each pair of diagrams, 
the interaction and electron lines are $U$ and $R_{\sigma}({i}\varepsilon_l)$, respectively, for $\Sigma_\sigma({i}\varepsilon_l)$ of the Hubbard model, while they are
$\tilde{U}$ and $\tilde{G}_{\sigma}({i}\varepsilon_l)$, respectively, for $\tilde{\Sigma}_\sigma({i}\varepsilon_l)$ of the Anderson model.
Provided that a couple of
\begin{subequations}\label{EqMap0}
\begin{align}\label{EqMap1}
\tilde{U}= U, 
\end{align}
and
\begin{align}\label{EqMap2}
\tilde{G}_{\sigma}({i}\varepsilon_l)=
R_{\sigma}({i}\varepsilon_l),
\end{align}
are satisfied, the contributions of any pair of diagrams are exactly the same as each other; thus, the mapping of Eq.\hskip2pt(\ref{EqMapS}) is satisfied.
The condition of Eq.\hskip2pt(\ref{EqMap2}) is equivalent to a couple of
\begin{align}\label{EqMap3}
\tilde{\mu}- \tilde{\epsilon}_d=\mu-\epsilon_d,
\end{align}
where $\epsilon_d$ is the band center and $\epsilon_d=0$, and
\begin{align}\label{EqMap4}
\tilde{\Delta}(\varepsilon) &=
{\rm Im}\left[
\tilde{\Sigma}_\sigma(\varepsilon+{i}0)
+1/R_{\sigma}(\varepsilon+{i}0) \right] ,
\end{align} 
\end{subequations}
where $\tilde{\Sigma}_\sigma(\varepsilon+{i}0)$ and $R_{\sigma}(\varepsilon+{i}0)$ are the analytical continuations of $\tilde{\Sigma}_\sigma({i}\varepsilon_l )$ and $R_{\sigma}({i}\varepsilon_l)$, respectively, from the upper-half complex plane onto the real axis.
Since Eqs.\hskip2pt(\ref{EqMap1}) and (\ref{EqMap3}) are trivial,
Eq.\hskip2pt(\ref{EqMap4}) is a practical mapping condition for determining the mapped Anderson model.%
\footnote{Following the previous paper,\cite{toyama} it can be proved on the basis of Eq.\hskip2pt(\ref{EqMap4}) that $\tilde{\Delta}(\varepsilon) \ge 0$ for any possible self-consistent solution, as anticipated.}

If a series of Feynman diagrams, single-site or multisite ones, for the Hubbard model are divergent in the thermodynamic limit of $L\rightarrow+\infty$ followed by $T\rightarrow 0\hskip2pt$K,
the limit has to be taken at this stage.
The mapped Anderson model depends on $L$ and $T$. In the following part of this paper, we consider the mapped Anderson model in the limit of $L\rightarrow+\infty$, if necessary, followed by $T\rightarrow 0\hskip2pt$K.

In the wave-number representation, the self-energy is given by
\begin{align}\label{EqSigmaDecomposedK}
\Sigma_\sigma({i}\varepsilon_l, {\bm k}) =
\tilde{\Sigma}_\sigma({i}\varepsilon_l)
+ \Delta\Sigma_\sigma({i}\varepsilon_l, {\bm k}),
\end{align}
where Eq.\hskip2pt(\ref{EqMapS}) is used for the single-site self-energy, and
\begin{align}
\Delta\Sigma_\sigma({i}\varepsilon_l, {\bm k}) &=
\frac1{L}\sum_{ij}e^{-i{\bm k}\cdot({\bm x}_i-{\bm x}_j)}\Delta\Sigma_{ij;\sigma}({i}\varepsilon_l).
\end{align}
The Green function is given by
\begin{align}\label{EqGreen}
G_\sigma({i}\varepsilon_l, {\bm k}) &
= \frac1{L}\sum_{ij}e^{-i{\bm k}\cdot({\bm x}_i-{\bm x}_j)}R_{ij;\sigma}({i}\varepsilon_l)
= \frac1{\displaystyle {i}\varepsilon_l  +  \mu - E({\bm k})  
- \Sigma_{\sigma}({i}\varepsilon_l, {\bm k})}.
\end{align}
%
%
The site-diagonal Green function is given by
\begin{align}\label{EqGreenR}
R_\sigma({i}\varepsilon_l) &
= \frac1{L}\sum_{\bm k}\frac1{\displaystyle {i}\varepsilon_l  +  \mu - E({\bm k})  
- \Sigma_{\sigma}({i}\varepsilon_l, {\bm k})}.
\end{align}
The density of states is simply given by
\begin{subequations}\label{EqDOS}
\begin{align}\label{EqDOS1}
\rho(\varepsilon)  
= - \frac1{\pi}{\rm Im}\hskip1pt R_\sigma(\varepsilon+{i}0).
\end{align}
Because of Eq.\hskip2pt(\ref{EqMap2}),
it is equal to that of the mapped Anderson model either under or beyond S$^3$A:
\begin{align}\label{EqDOS2}
\rho(\varepsilon) & 
= - \frac1{\pi}{\rm Im}\hskip1pt R_\sigma(\varepsilon+{i}0)
= -\frac1{\pi}{\rm Im} \tilde{G}_{\sigma}(\varepsilon+{i}0).
\end{align}
\end{subequations}
This equality is equivalent to the mapping condition and is crucial in this paper.





When $\mu = U/2$, not only the Hubbard model but also the mapped Anderson model is half-filled and symmetrical:
\begin{align}\label{EqMuEdU}
\mu = \tilde{\mu} - \tilde{\epsilon}_d = U/2 =\tilde{U}/2, \quad
\langle {\cal N}\rangle/L =\langle \tilde{n}_{\uparrow}+ \tilde{n}_{\downarrow}\rangle=1,
\end{align}
where ${\cal N} = \sum_{i\sigma}n_{i\sigma}$ and $\left<\cdots\right>$ stands for the statistical average.
Since the particle-hole symmetry exists in any half-filled symmetrical model, various equalities are inevitably satisfied: In particular,
\begin{align}\label{EqP-H}
{\rm Re}\tilde{\Sigma}_\sigma(+{i}0)-\mu=0, \quad
{\rm Re}\tilde{\Gamma}(+{i}0)=0, \quad
E({\bm k}_{\rm F})+{\rm Re}\Sigma_\sigma(+{i}0,{\bm k}_{\rm F})-\mu=0, 
\end{align}
are crucial,
where ${\bm k}_{\rm F}$ is defined by Eq.\hskip2pt(\ref{EqHyper-surface}) or $E({\bm k}_{\rm F})=0$.

It is easy to see from Eqs.\hskip2pt(\ref{EqGreenAM}), (\ref{EqGreenR}), (\ref{EqDOS}), and (\ref{EqP-H}) that, unless ${\rm Im}\bigl[\tilde{\Sigma}_\sigma(\varepsilon+{i}0)+\tilde{\Gamma}(\varepsilon+{i}0)\bigr]$ is divergent at $\varepsilon=0$, then $\rho(0)>0$ and not $\rho(0)=0$ in the Anderson model and that, unless ${\rm Im}\Sigma_\sigma(\varepsilon+{i}0,{\bm k}_{\rm F})$ is divergent at $\varepsilon=0$, then $\rho(0)>0$ and not $\rho(0)=0$ in the Hubbard model.
A couple of these features are also crucial in this paper.

\subsection{Kondo effect}
\label{SecKondoEffect}
\subsubsection{Normal Fermi liquid in the Anderson model}
\label{SecNFL-AM}
The Kondo problem has already been solved;\cite{yosida,poorman,wilsonKG,nozieres,yamada1,yamada2,shiba}
in particular, the Bethe-ansatz solution was given to the Anderson model in which $\tilde{\Delta}(\varepsilon)$ is constant as well as to the $s$-$d$ model in which the density of states of conduction electrons is constant.\cite{exact1,exact2,exact3,exact4}
For example, we consider the Anderson model defined by Eq.\hskip2pt(\ref{EqAnderson}).
The static spin susceptibility of localized electrons is given by
\begin{align}\label{EqChiAM}
\tilde{\chi}_s(T) =
(1/2)g^2\mu_{\rm B}^2 
\left[\tilde{\chi}_{\uparrow\uparrow}(T)- \tilde{\chi}_{\uparrow\downarrow}(T)\right],
\end{align}
where $g$ and $\mu_{\rm B}$ are the $g$ factor and Bohr magneton, respectively, of localized electrons, and
\begin{align}\label{EqSusAM1}
\tilde{\chi}_{\sigma\sigma^\prime}(T) &
= 
\int_{0}^{1/k_{\rm B}T} 
\hskip-3pt d\tau \left<e^{\tau(\tilde{\cal H}-\tilde{\mu}\tilde{\cal N}))}\tilde{n}_{d\sigma} e^{-\tau(\tilde{\cal H}-\tilde{\mu}\tilde{\cal N})} \tilde{n}_{d\sigma^\prime}\right>,
\end{align}
where $\tilde{\cal N} =\sum_{\sigma}\tilde{n}_{d\sigma} +\sum_{{\bm k}\sigma} \tilde{c}_{{\bm k}\sigma}^\dag \tilde{c}_{{\bm k}\sigma}$.
Since the most crucial physics of the Kondo effect is the quenching of a local moment by conduction electrons,\cite{yosida,poorman,wilsonKG}
determining whether the Fermi surface of conduction electrons exists is crucial.
The Fermi surface is defined by $\tilde{\mu}=\tilde{E}_c({\bm k})$. 
If the Fermi surface exists and $|\tilde{V}_{\bm k}|^2$ is nonzero at least on a part of the Fermi surface, it follows that 
\begin{align}\label{EqPositiveDelta}
0<\tilde{\Delta}(\varepsilon)<+\infty,
\end{align}
at least for $\varepsilon\simeq 0$, so that the ground state is a singlet and a normal Fermi liquid.
The Kondo temperature $T_{\rm K}$ is defined by%
\footnote{
This definition of $T_{\rm K}$ is different from Wilson's\cite{wilsonKG} by a numerical factor, even in the $s$-$d$ limit.}
%
\begin{align}\label{EqDefTK} 
k_{\rm B}T_{\rm K} =
(1/4)g^2\mu_{\rm B}^2/\tilde{\chi}_s(0).
\end{align}
%
If $T\ll T_{\rm K}$, $\tilde{\chi}_s(T)$ exhibits the Pauli paramagnetism; if $T\gg T_{\rm K}$, $\tilde{\chi}_s(T)$ obeys the Curie-Weiss law, i.e., localized electrons behave like a local moment or a free spin.

If $\tilde{\Delta}(0)$ is neither zero nor infinite and if $T_{\rm K}$ is nonzero, the ground state is a normal Fermi liquid, so that
the self-energy of localized electrons is analytical at $\varepsilon=0$ and it can be expanded in a way such that\cite{yamada1,yamada2,shiba}
\begin{align}\label{EqExpansionAM}
\tilde{\Sigma}_\sigma(\varepsilon + {i}0) &=
\frac1{2}U + \bigl(1 -\tilde{\phi}_1 \bigr) \varepsilon
- i \tilde{\phi}_2 \varepsilon^2/\tilde{\Delta}(0)
- i \tilde{\phi}_3 \left(k_{\rm B}T\right)^2/\tilde{\Delta}(0)
+ \cdots,
\end{align}
for $|\varepsilon|\ll k_{\rm B}T_{\rm K}$ and $T\ll T_{\rm K}$, 
where $\tilde{\phi}_1 > 1$, $\tilde{\phi}_2 >0$, and $\tilde{\phi}_3 >0$.
When $T=0\hskip2pt$K, according to Eqs.\hskip2pt(\ref{EqGreenAM}), (\ref{EqDOS}), (\ref{EqP-H}), and (\ref{EqExpansionAM}), the density of states of localized electrons at the chemical potential is simply given by
\begin{align}\label{EqDOS-AM}
\rho(0) = 1/\bigl[\pi\tilde{\Delta}(0)\bigr].
\end{align}
If $\tilde{\Delta}(\varepsilon)$ is constant such that $\tilde{\Delta}(\varepsilon)=\tilde{\Delta}(0)$, Anderson's compensation theorem is valid;\cite{compensation} there is no additional contribution of conduction electrons when either $\tilde{V}_{\bm k}$ or $\tilde{U}$ is introduced. Then, the expansion coefficients are simply given by\cite{yamada1,yamada2,shiba}
%
%
\begin{align}\label{EqExpansionAM1}
\tilde{\phi}_1=\tilde{\chi}_{\uparrow\uparrow}(0)
\bigl[\pi\tilde{\Delta}(0)\bigr],
\quad
\tilde{\phi}_2=\tilde{\phi}_3=\left\{\tilde{\chi}_{\uparrow\downarrow}(0)
\bigl[\pi\tilde{\Delta}(0)\bigr]\right\}^2\hskip-2pt\big/2.
\end{align}
The specific-heat coefficient of localized electrons is given by 
\begin{align}\label{EqHeatAM}
\tilde{\gamma} &=
(2/3)\pi^2k_{\rm B}^2\tilde{\phi}_1/\bigl[\pi\tilde{\Delta}(0)\bigr].
\end{align}
The Fermi-liquid relation above is consistent with the Bethe-ansatz solution.\cite{exact1,exact2,exact3,exact4}

If $\tilde{\Delta}(\varepsilon)$ depends on $\varepsilon$, the compensation theorem is not valid, in general; there are additional contributions of conduction electrons.
If the compensation theorem is valid or not valid,
the static spin susceptibility and specific-heat coefficient of localized electrons, which include no additional contribution of conduction electrons, are given by Eqs.\hskip2pt(\ref{EqChiAM}) and (\ref{EqHeatAM}), respectively,
but with $\tilde{\phi}_1\ne \tilde{\chi}_{\uparrow\uparrow}(0)\bigl[\pi\tilde{\Delta}(0)\bigr]$ in general.
The Wilson ratio is defined by
\begin{align}\label{EqWilsonRatio}
\tilde{W}_s 
=\frac{\tilde{\chi}_s(0)}{\tilde{\gamma}}
\frac{(2/3)\pi^2k_{\rm B}^2}{(1/2)g^2\mu_{\rm B}^2},
\end{align}
which is of the order of unity or $O(1)$, at least, when the onsite $\tilde{U}$ is repulsive.

If $T_{\rm K}$ is nonzero, the ground state of the Anderson model is a normal Fermi liquid. 
When $T=0\hskip2pt$K,
${\rm Im}\tilde{\Sigma}_\sigma(+{i}0) =0$ and the entropy or residual entropy of localized electrons is zero.
When $0\hskip2pt{\rm K}<T\ll T_{\rm K}$, ${\rm Im}\tilde{\Sigma}_\sigma(+{i}0)$ is negatively nonzero and small, and the entropy of localized electrons is much smaller than $k_{\rm B}\ln 2$.
When $T\ll T_{\rm K}$, localized electrons never behave like a free spin. 
When $T\gtrsim T_{\rm K}$, ${\rm Im}\tilde{\Sigma}_\sigma(+{i}0)$ is negatively large, the entropy of localized electrons is as large as $k_{\rm B}\ln 2$, and localized electrons behave like a free spin. 
This crossover as a function of $T$ is crucial.\cite{wilsonKG}


\subsubsection{Possible gapped phases in the Anderson model}
\label{SecGappedPhaseAM}
If Eq.\hskip2pt(\ref{EqPositiveDelta}) is not satisfied, it is possible that $T_{\rm K}=0\hskip2pt$K or $T_{\rm K}=+0\hskip2pt$K.
If $T_{\rm K}=0\hskip2pt$K or $T_{\rm K}=+0\hskip2pt$K, 
the self-energy $\tilde{\Sigma}_\sigma(\varepsilon+{i}0)$ is singular at $\varepsilon=0$.
We examine how singular $\tilde{\Sigma}_\sigma(\varepsilon+{i}0)$ is in the case of $T_{\rm K}=0\hskip2pt$K or $T_{\rm K}=+0\hskip2pt$K.

According to Hubbard's theory,\cite{Hubbard1}
in the so-called atomic limit of $W/U\rightarrow 0$, the self-energy  of the Hubbard model is given by 
\begin{align}\label{EqSSA-S1}
\Sigma_\sigma(\varepsilon+{i}0)
= \frac1{2} U+ \frac{U^2}{4}\frac1{\varepsilon+{i}0}  .
\end{align}
Since Hubbard's theory is under SSA,
it is anticipated on the basis of S$^3$A or the Kondo-lattice theory that $\Sigma_\sigma(\varepsilon+{i}0)$ given by Eq.\hskip2pt(\ref{EqSSA-S1}) can be approximately used for $\tilde{\Sigma}_\sigma(\varepsilon + {i}0)$ of the Anderson model.
The atomic limit in the Hubbard model corresponds to the limit of $T_{\rm K}\rightarrow0\hskip2pt$K or the case of $T_{\rm K}=0\hskip2pt$K in the Anderson model. 
It is anticipated that, if $T_{\rm K}=+0\hskip2pt$K or $T_{\rm K}=0\hskip2pt$K, the self-energy $\tilde{\Sigma}_\sigma(\varepsilon+{i}0)$ of the Anderson model has a pole at $\varepsilon=0$.

Next, we examine how the self-energy $\tilde{\Sigma}_\sigma(\varepsilon + {i}0)$ of the Anderson model becomes singular as $T_{\rm K}\rightarrow 0\hskip2pt$K.
For example, we consider a function defined by
\begin{align}\label{EqFuncSigma}
\tilde{\Sigma}_\sigma(\varepsilon+{i}0; \gamma_0)&
= \frac1{2}U + \frac{U^2}{4}
\int_{+0}^{+\infty} \frac{d\gamma}{\gamma_0}\int_{+0}^{\pi/2}d\theta \hskip2pt \zeta \hskip-2pt \left(\gamma/\gamma_0, \theta\right) \Xi(\varepsilon; \gamma, \theta),
\end{align}
where $\gamma_0\simeq k_{\rm B}T_{\rm K}$ is a parameter, $\zeta(x,\theta)$ is a functional parameter, and
\begin{align}\label{EqFuncXi}
\Xi(\varepsilon; \gamma, \theta)=
\frac1{2\cos\theta} 
\left[
\frac{-e^{-i(\pi-\theta)}}{\varepsilon-\gamma e^{-i(\pi-\theta)}}
+ \frac{-e^{-i\theta}}{\varepsilon-\gamma e^{-i\theta}}
\right]
=
\frac{-\varepsilon(\gamma^2-\varepsilon^2)- i2\varepsilon^2\gamma\sin\theta }{\varepsilon^4 -2 \varepsilon^2\gamma^2 \cos2\theta+\gamma^4}.
\end{align}
Here, we assume that $\zeta (x, \theta)$ satisfies
$\zeta (x,\theta)\ge 0$,  $\zeta(x\rightarrow +\infty, \theta)= 0$, and
\begin{align}
\int_{+0}^{+\infty}\hskip-5pt dx\int_{+0}^{\pi/2}d\theta \hskip2pt \zeta(x,\theta) =
\int_{+0}^{+\infty}\hskip-5pt dx \int_{+0}^{\pi/2}d\theta\hskip2pt \zeta(x,\theta) x =1.
\end{align}
It is easy to see that $\Xi(\varepsilon; \gamma, \theta)$ as a function of $\varepsilon$ is analytical in the upper-half complex plane at least for  $\gamma>0$ and $0<\theta<\pi/2$.
It should be noted that
\begin{align}\label{EqDeltaVar}
\frac1{\pi} \int_{-\infty}^{+\infty}d\varepsilon
\frac{2\gamma\varepsilon^2\sin\theta}{\varepsilon^4-2 \varepsilon^2\gamma^2\cos2\theta+\gamma^4} =1. 
\end{align}
Thus, $-(1/\pi){\rm Im}\Xi(\varepsilon;\gamma\rightarrow 0,\theta)$ is the delta function $\delta(\varepsilon)$ or, at least, it can be treated as the delta function, although ${\rm Im}\Xi(\varepsilon=0;\gamma,\theta)=0$ for any nonzero $\gamma$.
It follows that
\begin{subequations}
\begin{align}\label{EqSTG1}
&\lim_{\varepsilon/\gamma_0\rightarrow 0}
\tilde{\Sigma}_\sigma(\varepsilon+{i}0; \gamma_0) 
= \frac1{2}U - \frac{U^2}{4\gamma_0^2}
\left(\tilde{s}_1 \varepsilon + {i} \tilde{s}_2\frac{\varepsilon^2}{\gamma_0} + \cdots\right), 
\\ \label{EqSTG2} 
& \lim_{\varepsilon/\gamma_0\rightarrow \pm\infty}
\tilde{\Sigma}_\sigma(\varepsilon+{i}0; \gamma_0) 
=\frac1{2}U + \frac{U^2}{4}\left( \frac1{\varepsilon} -i \tilde{s}_3 \hskip1pt\frac{\gamma_0}{\varepsilon^2}
+ \cdots \right), 
\end{align}
\end{subequations}
where
\begin{subequations}
\begin{align}
\tilde{s}_1 =  
\int_{+0}^{+\infty} \hskip-5pt dx \int_{+0}^{\pi/2}d\theta\hskip2pt \zeta(x, \theta)\frac1{x^2}, & \quad 
\tilde{s}_2 =
\int_{+0}^{+\infty} \hskip-5pt dx \int_{+0}^{\pi/2}d\theta\hskip2pt \zeta(x,\theta) \frac{2\sin\theta}{x^3},
\\ & \hskip-60pt
\tilde{s}_3=  \int_{+0}^{+\infty} \hskip-5pt dx \int_{+0}^{\pi/2}d\theta\hskip2pt \zeta(x,\theta) 2x\sin\theta.
\end{align}
\end{subequations}
If $\gamma_0 \simeq k_{\rm B}T_{\rm K}$ and $\zeta(x, \theta)$ are properly chosen, 
$\tilde{\Sigma}_\sigma(\varepsilon+{i}0; \gamma_0)$ behaves in almost the same way as $\tilde{\Sigma}_\sigma(\varepsilon+{i}0)$ given by Eq.\hskip2pt(\ref{EqExpansionAM}) for $|\varepsilon|\ll \gamma_0$ or $|\varepsilon|\ll k_{\rm B}T_{\rm K}$.
On the other hand, it follows from Eqs.\hskip2pt(\ref{EqDeltaVar}) and (\ref{EqSTG2}) that 
\begin{align}
\lim_{\gamma_0\rightarrow 0}\lim_{\varepsilon/\gamma_0\rightarrow \pm\infty}
\tilde{\Sigma}_\sigma(\varepsilon+{i}0; \gamma_0) &
= \frac1{2}U + \frac{U^2}{4} \frac1{\varepsilon+{i}0}.
\end{align} 
Thus, 
$\tilde{\Sigma}_\sigma(\varepsilon+{i}0; \gamma_0\rightarrow 0)$ has a pole or a variation of the pole at $\varepsilon=0$; the pole is not simple if $\gamma_0$ is nonzero.
It is anticipated that the analytical self-energy becomes singular in
this way or a similar way as $\gamma_0\rightarrow 0\hskip1pt$ or $T_{\rm K}\rightarrow 0\hskip2pt$K.

It is anticipated on the basis of the arguments above that there is no essential difference between the self-energy characterized by $T_{\rm K}=0\hskip2pt$K and that by $T_{\rm K}=+0\hskip2pt$K. 
If the self-energy $\tilde{\Sigma}_\sigma(\varepsilon+{i}0)$ has a pole or the variation of the pole discussed above at $\varepsilon=0$, if the residue of the pole is large enough, if ${\rm Im} \tilde{\Sigma}_\sigma(\varepsilon+{i}0)=0$ for $|\varepsilon|<\epsilon_{\rm g}/2$ but $\varepsilon=0$, and if $\tilde{\Delta}(\varepsilon)=0$ for $|\varepsilon|<\epsilon_{\rm g}/2$, a gap as large as $\epsilon_{\rm g}$ opens in the Anderson model.%
\footnote{According to Hubbard's theory,\cite{Hubbard1,Hubbard2} when $U/W$ is so large that a Hubbard gap as large as $\epsilon_{\rm g}$ opens, $\Sigma_\sigma(\varepsilon+{i}0)$ has a pole at $\varepsilon=0$, the residue of the pole is large enough, and ${\rm Im}\Sigma_\sigma(\varepsilon+{i}0)=0$ for $|\varepsilon|<\epsilon_{\rm g}/2$ but $\varepsilon=0$, as in the Anderson model.}

To be precise, it has never been fully clarified what singularity the self-energy $\tilde{\Sigma}_\sigma(\varepsilon+{i}0)$ of the Anderson model has to have at $\varepsilon=0$ in the case of $T_{\rm K}=0\hskip2pt$K and $T_{\rm K}=+0\hskip2pt$K, i.e., a simple pole, a variation of the pole, or some thing else.
However, at least, it is certain that, unless $\tilde{\Delta}(0)$ is exactly zero or infinite and unless $T_{\rm K}$ is the absolute zero or infinitesimal, the self-energy $\tilde{\Sigma}_\sigma(\varepsilon+{i}0)$ has no type of singularity at $\varepsilon=0$ and it can be expanded as in Eq.\hskip2pt(\ref{EqExpansionAM}).

%

Unless $\tilde{\Delta}(0)$ is infinite,%
\footnote{If $\tilde{\Delta}(0)$ is divergent, a gap inevitably opens.}
a gap can open only if at least ${\rm Im}\tilde{\Sigma}_\sigma(\varepsilon+{i}0)$ is divergent at $\varepsilon=0$. 
The divergence of ${\rm Im}\tilde{\Sigma}_\sigma(\varepsilon+{i}0)$ at $\varepsilon=0$ is possible only if $T_{\rm K}=0\hskip2pt$K or $T_{\rm K}=+0\hskip2pt$K.
If $T_{\rm K}=0\hskip2pt$K, the residual entropy is nonzero; localized electrons behave like a free spin even at $T=0\hskip2pt$K.
If $T=+0\hskip2pt$K, the entropy is zero at $T=0\hskip2pt$K but is nonzero at $T=+0\hskip2pt$K; localized electrons behave like a free spin even at $T=+0\hskip2pt$K.
The third law of thermodynamics is broken in any gapped phase of the Anderson model characterized by $T_{\rm K}=0\hskip2pt$K or $T_{\rm K}=+0\hskip2pt$K.%
\footnote{
One of the expressions of the third law of thermodynamics is that the residual entropy has to be zero.
Another expression is that the absolute zero Kelvin or $T=0\hskip2pt$K can never be reached.
Since $T=0\hskip2pt$K can never be reached, it is possible to extend the first expression to the third one: Entropy has to be continuously decreasing and vanishing as $T\rightarrow0\hskip2pt$K.
If the possibility of an ordered ground state in an actual system in three dimensions is considered, the third expression of the third law can be extended to the fourth one: Entropy has to be continuously decreasing and vanishing faster than linearly in $T$ as $T\rightarrow0\hskip2pt$K.}

\subsubsection{Kondo temperature of the mapped Anderson model}
According to the Kondo-lattice theory, the Kondo temperature $T_{\rm K}$ or $k_{\rm B}T_{\rm K}$ of the mapped Anderson model is also the energy scale of single-site quantum spin fluctuations in the Hubbard model.%
\footnote{Since the mapped Anderson model is determined for a given $T$, it depends on the given $T$, in general. Thus, the Kondo temperature $T_{\rm K}$ in the Hubbard model depends on $T$, in general.}
Thus, we also call the energy scale in the Hubbard model the Kondo temperature;
it is also denoted by $T_{\rm K}$ or $k_{\rm B}T_{\rm K}$. 
If the Kondo temperature $T_{\rm K}$ of the mapped Anderson model is the absolute zero or infinitesimal or if the self-energy $\tilde{\Sigma}_\sigma(\varepsilon+{i}0)$ of the mapped Anderson model is divergent at $\varepsilon=0$, it can be concluded that at least the single-site part of the residual entropy or entropy at $T=+0\hskip2pt$K is nonzero in the Hubbard model.

\section{Nongapped and gapped phases under S$^3$A}
\label{SecDMFT}
\subsection{Infinite bandwidth model}
\label{SecInfiniteW}
We consider under S$^3$A the infinite bandwidth model in $D\ge 1$ dimensions, in which the bare density of states $\rho_0(E)$ is given by Eq.\hskip2pt(\ref{EqDOS-Lorentzian}).
Since $\rho_0(E)$ is of a Lorentzian type, the problem is exactly the same as that for the Bethe-lattice model in $D\rightarrow+\infty$ dimensions,\cite{georges} which is exactly solvable.

The multisite $\Delta\Sigma_\sigma({i}\varepsilon_l, {\bm k})$ vanishes under S$^3$A. It follows from Eq.\hskip2pt(\ref{EqGreenR}) that
\begin{subequations}\label{EqDOS-SSSA}
\begin{align}\label{EqDOS-SSSA1}
R_\sigma(\varepsilon+{i}0) &
= \int_{-\infty}^{+\infty} \hskip-3pt dE \rho_0(E)
\frac1{\varepsilon+\mu -E -\tilde{\Sigma}_\sigma(\varepsilon+{i}0)}
\\ \label{EqDOS-SSSA2} &
= \frac1{\varepsilon+\mu  -\tilde{\Sigma}_\sigma(\varepsilon+{i}0)+i|t_\infty|}.
\end{align}
\end{subequations}
It follows from Eq.\hskip2pt(\ref{EqMap4}) that
%
$\tilde{\Delta}(\varepsilon) = |t_\infty|$.
%
The self-consistent Kondo problem is reduced to a {\it non-self-consistent} Kondo problem, to which the Bethe-ansatz solution was given.\cite{exact1,exact2,exact3,exact4}
The ground state is a normal Fermi liquid; 
$T_{\rm K}$ is nonzero.
The density of states is simply given by
$\rho(\varepsilon)=-(1/\pi){\rm Im}R_\sigma(\varepsilon+{i}0)$, where  
$R_\sigma(\varepsilon+{i}0)$ is given by Eq.\hskip2pt(\ref{EqDOS-SSSA2}).
%
%
Since the single-site $\tilde{\Sigma}_\sigma(\varepsilon+{i}0)$ can be expanded as shown in Eq.\hskip2pt(\ref{EqExpansionAM}), the density of states in the vicinity of the chemical potential is approximately given by that of a Lorentzian type, whose bandwidth is $O(k_{\rm B}T_{\rm K})$.
In particular, 
%
$\rho(0)= \rho_0(0)=1/(\pi |t_\infty|)$
%
at $T=0\hskip2pt$K.
When $T=0\hskip2pt$K, $\rho(0)$ is never renormalized by $U$.

\subsection{Finite bandwidth model}
\label{SecFiniteW}
When the bare bandwidth $W$ is finite and the onsite $U$ is large enough such that $U\gtrsim W$, a complete gap opens at $T=0\hskip2pt$K under DMFT, \cite{kotliar,moeller, bulla} which is simply S$^3$A.
In this subsection, we list all the possible types of solution under  S$^3$A and we examine the nature of each of them.
We do not specify a precise model for $\rho_0(E)$; we only assume that $\rho_0(E)>0$ for $|E|<W/2$ and $\rho_0(E)=0$ for $|E|>W/2$.

The density of states is given by Eq.\hskip2pt(\ref{EqDOS-SSSA1})
%
%
for the Hubbard model and
\begin{align}\label{EqDOS-DMFT-A}
\rho(\varepsilon) &
= - \frac1{\pi}{\rm Im} 
\frac1{\varepsilon +\mu -\tilde{\Sigma}_\sigma(\varepsilon+{i}0)-\tilde{\Gamma}(\varepsilon+{i}0)},
\end{align}
for the mapped Anderson model. 
Under S$^3$A, $\tilde{\Sigma}_\sigma(\varepsilon+{i}0)$ and $\tilde{\Gamma}(\varepsilon+{i}0)$ are self-consistently calculated or determined in a way such that Eqs.\hskip2pt(\ref{EqDOS-SSSA1}) and (\ref{EqDOS-DMFT-A}) are equal to each other.
According to Eq.\hskip2pt(\ref{EqHybridGamma0}),
if one of the real and imaginary parts of $\tilde{\Gamma}(\varepsilon+{i}0)$ is finite and continuous at $\varepsilon=\varepsilon_0$, the other is also finite and continuous at $\varepsilon=\varepsilon_0$; and
if one of them is discontinuous or divergent at $\varepsilon=\varepsilon_0$, the other is divergent or discontinuous at $\varepsilon=\varepsilon_0$.
In general, 
\begin{align}\label{EqSelfKR}
\tilde{\Sigma}_\sigma(\varepsilon+{i}0) -\mu &
= -\frac1{\pi}\int_{-\infty}^{+\infty} dx 
\frac{{\rm Im}\tilde{\Sigma}_\sigma(x+{i}0)}
{\varepsilon+{i}0-x},
\end{align}
for any symmetrical model.
According to Eq.\hskip2pt(\ref{EqSelfKR}), the real and imaginary parts of
$\tilde{\Sigma}_\sigma(\varepsilon+{i}0)$ satisfy the very same relation as that for $\tilde{\Gamma}(\varepsilon+{i}0)$ discussed above. 

First, we consider a possible solution in which ${\rm Re}\tilde{\Sigma}_\sigma(\varepsilon+{i}0)$ is finite and continuous at $\varepsilon=0$; 
${\rm Im} \tilde{\Sigma}_\sigma(\varepsilon+{i}0)$ is also finite and continuous at $\varepsilon=0$.
Since ${\rm Re}\tilde{\Sigma}_\sigma(+{i}0)-\mu=0$ and ${\rm Re}\tilde{\Gamma}(+{i}0)=0$ because of the particle-hole symmetry,
it follows from either Eq.\hskip2pt(\ref{EqDOS-SSSA1}) or (\ref{EqDOS-DMFT-A}) that $\rho(0)>0$. 
Since $\tilde{\Sigma}_\sigma(\varepsilon+{i}0)$ is finite and continuous at $\varepsilon=0$, according to the mapping condition of Eq.\hskip2pt(\ref{EqMap4}), $\tilde{\Delta}(\varepsilon)=(-1/\pi){\rm Im}\tilde{\Gamma}(\varepsilon+{i}0)$ is nonzero, finite, and continuous at $\varepsilon=0$; 
${\rm Re}\tilde{\Gamma}(\varepsilon+{i}0)$ is also finite and continuous at $\varepsilon=0$.
Then, $\rho(\varepsilon)>0$ for $\varepsilon\simeq 0$. 
%
Neither possibility, ${\rm Im} \tilde{\Sigma}_\sigma(+{i}0)<0$ nor ${\rm Im} \tilde{\Sigma}_\sigma(+{i}0)=0$, can be excluded. 
According to the correspondence to the mapped Anderson model,
any solution characterized by ${\rm Im} \tilde{\Sigma}_\sigma(+{i}0)=0$ is one for a normal Fermi liquid at $T=0\hskip2pt$K and
any solution characterized by ${\rm Im} \tilde{\Sigma}_\sigma(+{i}0)<0$ is one for a normal Fermi liquid at $T>0\hskip2pt$K.
If ${\rm Im} \tilde{\Sigma}_\sigma(+{i}0)=0$, it follows from Eq.\hskip2pt(\ref{EqDOS-SSSA1}) that $\rho(0) = \rho_0(0)$ for the Hubbard model and it follows from Eq.\hskip2pt(\ref{EqDOS-DMFT-A}) that $\rho(0)=1/[\pi\tilde{\Delta}(0)]$ for the Anderson model, as shown by Eq.\hskip2pt(\ref{EqDOS-AM}); thus, $\tilde{\Delta}(0)=1/[\pi\rho_0(0)]$.
Neither $\rho(0)$ nor $\tilde{\Delta}(0)$ is renormalized by $U$. 
As discussed in the following paragraphs, a gap inevitably opens if ${\rm Re}\tilde{\Sigma}_\sigma(\varepsilon+{i}0)$ is discontinuous or divergent at $\varepsilon=0$.
Thus, the possible solution examined in this paragraph is the only possible one for a nongapped phase under S$^3$A, so that
it is none other than the metallic solution obtained under DMFT for a sufficiently small $U$, i.e., $U\lesssim W$.\cite{kotliar,moeller,bulla}
This characterization is never inconsistent with the feature of the metallic phase under DMFT,\cite{kotliar,moeller,bulla} or rather consistent with it; for example, the constancy of $\rho(0) = \rho_0(0)$ and $\tilde{\Delta}(0)=1/[\pi\rho_0(0)]$ at $T=0\hskip2pt$K as a function of $U$ is realized in DMFT.\cite{kotliar,moeller,bulla}

Second, we consider one in which
${\rm Re}\tilde{\Sigma}_\sigma(\varepsilon+{i}0)$ remains finite as $\varepsilon\rightarrow \pm 0$
but it is discontinuous at $\varepsilon=0$. 
It follows from Eq.\hskip2pt(\ref{EqSelfKR}) that
\begin{align}\label{EqImaginaryDiv}
\lim_{\varepsilon\rightarrow \pm 0}{\rm Im}\tilde{\Sigma}_\sigma(\varepsilon+{i}0) = - \infty.
\end{align}
Since ${\rm Im}\tilde{\Sigma}_\sigma(\varepsilon+{i}0)$ is nonzero at least for a sufficiently small $|\varepsilon|$, if this type of solution is possible, no complete gap but only a pseudo-gap or a zerogap opens. 

Third, we consider one in which ${\rm Re}\tilde{\Sigma}_\sigma(\varepsilon+{i}0)$ as well as ${\rm Im}\tilde{\Sigma}_\sigma(\varepsilon+{i}0)$ continuously diverges as $\varepsilon\rightarrow 0$: 
\begin{align}
\lim_{\varepsilon\rightarrow - 0}{\rm Re}\tilde{\Sigma}_\sigma(\varepsilon+{i}0) & = - \infty, \qquad
\lim_{\varepsilon\rightarrow + 0}{\rm Re}\tilde{\Sigma}_\sigma(\varepsilon+{i}0) = + \infty, 
\end{align}
as well as Eq.\hskip2pt(\ref{EqImaginaryDiv}).
If this type of solution is possible, a zerogap opens in it.

Fourth, we consider one in which
${\rm Re}\tilde{\Sigma}_\sigma(\varepsilon+{i}0)$ discontinuously  diverges at $\varepsilon=0$, i.e., ${\rm Re}\tilde{\Sigma}_\sigma(\varepsilon+{i}0)$ includes the delta function $\delta(\varepsilon)$.
In this case, $\tilde{\Sigma}_\sigma(\varepsilon+{i}0)$ cannot be analytical in the upper-half plane.
No self-consistent solution of this type is possible.

Last, we consider a possible solution in which
${\rm Re}\tilde{\Sigma}_\sigma(\varepsilon+{i}0)$ continuously diverges as $\varepsilon\rightarrow \pm 0$ and
${\rm Im}\tilde{\Sigma}_\sigma(\varepsilon+{i}0)$ discontinuously diverges at $\varepsilon=0$, 
i.e., $\tilde{\Sigma}_\sigma(\varepsilon+{i}0)$ has a pole at $\varepsilon=0$ such that
\begin{align}
\tilde{\Sigma}_\sigma(\varepsilon+{i}0) -\mu =
\frac{\tilde{\lambda}_\Sigma}{\varepsilon+{i}0} + \cdots.
\end{align}
Then, $\rho(0)=0$ in this type of solution.
A necessary condition for a complete gap as large as $\epsilon_{\rm g}$ to open in the Hubbard model is that $\tilde{\Sigma}_\sigma(\varepsilon+{i}0)$ has a pole at $\varepsilon=0$, the residue of the pole is large enough, as discussed below, and
${\rm Im}\tilde{\Sigma}_\sigma(\varepsilon+{i}0)=0$ for $|\varepsilon|<\epsilon_{\rm g}/2$ but $\varepsilon=0$. 
%
There is a zero point $z_E$ such that it satisfies
\begin{align}\label{EqZeroDMFT}
\varepsilon + \mu - z_E - {\rm Re}\tilde{\Sigma}_\sigma(\varepsilon+{i}0)=0,
\end{align}
for any finite $\varepsilon$ but $\varepsilon=0$.
Since $\rho_0(E)>0$ for $|E|<W/2$ and $\rho_0(E)=0$ for $|E|>W/2$,
$|\varepsilon + \mu - {\rm Re}\tilde{\Sigma}_\sigma(\varepsilon+{i}0)|>W/2$ for any $\varepsilon$ of $|\varepsilon|<\epsilon_{\rm g}/2$ has to be satisfied for a complete gap as large as $\epsilon_{\rm g}$ to open. Thus, the bandwidth $W$ has to be finite and the residue $\tilde{\lambda}_\Sigma$ has to be as large as\footnote{%
If the bare bandwidth is infinite such as $W=+\infty$, it is easy to see from Eqs.\hskip2pt(\ref{EqDOS-SSSA1}) and (\ref{EqZeroDMFT}) that no complete gap but only a zerogap can open. }
%
\begin{align}\label{EqLargeREsidue1}
\tilde{\lambda}_\Sigma > \frac1{4}\left(\epsilon_{\rm g}^2 + \epsilon_{\rm g}W\right).
\end{align}
If the bandwidth $2W$ is finite,
a possibility that a complete gap  opens can never be excluded.
This type of solution is the only possible one for a complete-gapped phase under S$^3$A, so that it is none other than the insulating solution obtained under DMFT for a sufficiently large $U$, i.e., $U\gtrsim W$.\cite{kotliar,moeller,bulla}
This characterization is not inconsistent with the feature of the gapped phase under DMFT,\cite{kotliar,moeller,bulla} or rather consistent with it, as discussed below.

In this subsection, we consider an unspecified finite bandwidth model.
Any possibility of nongapped, zerogapped, and complete-gapped phases cannot be excluded within the qualitative analysis of this section. 
However, it is certain that ${\rm Im}\tilde{\Sigma}_\sigma(+{i}0)$ is divergent in any possible type of gapped phase.
Thus, it can be concluded on the basis of the analysis in \S\ref{SecGappedPhaseAM} that $T_{\rm K}=0\hskip2pt$K or $T_{\rm K}=+0\hskip2pt$K in any gapped phase and  
that the residual entropy or entropy at $T=+0\hskip2pt$K per unit cell is therefore nonzero in it.
As regards a complete-gapped phase, in particular, the entropy at $T=+0\hskip2pt$K is $k_{\rm B}\ln 2$ per unit cell, as studied in Appendix\hskip2pt\ref{SecResidualEntropy}.

In any possible solution of a zerogap or complete-gap phase, inevitably $\tilde{\Delta}(0)=0$ under S$^3$A, as studied in Appendix\hskip2pt\ref{SecAnomaly}, as well as $\rho(0)=0$.
The feature that $\rho(0)=0$ and $\tilde{\Delta}(0)=0$ in the possible gapped phase is different from the feature that either $\rho(0)=\rho_0(0)$ or $\tilde{\Delta}(0)=1/[\pi\rho_0(0)]$ is nonzero and constant in the possible nongapped phase; therefore,
$\pi \tilde{\Delta}(0)\rho(0)=1$ in it.
If a metal-insulator transition occurs at $T=0\hskip2pt$K or $T=+0\hskip2pt$K under S$^3$A, not only the residual entropy or entropy at $T=+0\hskip2pt$K, but also $\rho(0)$ and $\tilde{\Delta}(0)$ jump at the transition. 
The jumps of $\rho(0)$ and $\tilde{\Delta}(0)$ are consistent with those in the metal-insulator transition in DMFT.\cite{kotliar,moeller,bulla}

\section{Gapped phase beyond S$^3$A}
\label{SecKLT}
%
In this section, we consider only possible complete-gapped phases at $T=0\hskip2pt$K or $T=+0\hskip2pt$K beyond S$^3$A in a finite or infinite bandwidth model.
A precise model for $\rho_0(\varepsilon)$ is neither specified.

The single-site Green function of the Hubbard model is given by $R_\sigma(\varepsilon+{i}0)$ defined by Eq.\hskip2pt(\ref{EqGreenR})
and the local Green function of the mapped Anderson model is given by Eq.\hskip2pt(\ref{EqGreenAM}), i.e., by
\begin{align}\label{EqR-AM}
\tilde{G}_\sigma(\varepsilon+{i}0) &=
\frac1{\varepsilon +{i}0 -\tilde{S}(\varepsilon+{i}0)},
\end{align}
where
\begin{align}\label{EqDOS-KLT-AS}
\tilde{S}(\varepsilon+{i}0) = 
\bigl[\tilde{\Sigma}_\sigma(\varepsilon+{i}0) -\mu\bigr]+\tilde{\Gamma}(\varepsilon+{i}0).
\end{align}
If the multisite $\Delta\Sigma_\sigma(\varepsilon+{i}0,{\bm k})$ is normal or anomalous, 
$\tilde{\Sigma}_\sigma(\varepsilon+{i}0)$ and $\tilde{\Gamma}(\varepsilon+{i}0)$, i.e., 
$\tilde{S}(\varepsilon+{i}0)$ of the mapped Anderson model is determined in a way such that $R_\sigma(\varepsilon+{i}0)$ and $\tilde{G}_\sigma(\varepsilon+{i}0)$ are equal to each other: $R_\sigma(\varepsilon+{i}0)=\tilde{G}_\sigma(\varepsilon+{i}0)$, as shown in Eq.\hskip2pt(\ref{EqMap2}).
In all possible solutions, it follows that
\begin{align}\label{EqKR-R}
R_\sigma(\varepsilon+{i}0) =
- \frac1{\pi}\int_{-\infty}^{+\infty} dx
\frac{\rho(x)}{\varepsilon+{i}0-x},
\end{align}
where 
$\rho(\varepsilon) = (- 1/\pi) {\rm Im} R_\sigma(\varepsilon +{i}0)$ is the density of states, and
\begin{align}\label{EqKR-S}
\tilde{S}(\varepsilon+i0) &=
- \frac1{\pi} \int_{-\infty}^{+\infty} dx \frac{{\rm Im} \tilde{S}(x+i0)}{\varepsilon+i0-x}. 
\end{align}

We assume that a complete gap as large as $\epsilon_{\rm g}$ opens in the $\rho(\varepsilon)$ of a self-consistent solution: $\rho(\varepsilon)=0$ for $|\varepsilon|<\epsilon_{\rm g}/2$. Then, Eq.\hskip2pt(\ref{EqKR-R}) can be described as
\begin{align}\label{EqRGap}
R_\sigma(\varepsilon+{i}0)  &=
- \frac1{\pi}\left(\int_{-\infty}^{-\epsilon_{\rm g}/2} dx+ \int_{+\epsilon_{\rm g}/2}^{+\infty} dx\right)
\frac{\rho(x)}{\varepsilon+{i}0-x}.
\end{align}
Then, for $|\varepsilon|<\epsilon_{\rm g}/2$, $R_\sigma(\varepsilon+{i}0)$ is real and  
$(d/d\varepsilon)R_\sigma(\varepsilon+{i}0)<0$.
Since ${\rm Re}R_\sigma(+{i}0)=0$ and ${\rm Re}\tilde{S}_\sigma(+{i}0)=0$ because of the particle-hole symmetry,
it is easy to see that $1/R_\sigma(\varepsilon+{i}0)$ has a pole at $\varepsilon=0$ and that, therefore, $\tilde{S}(\varepsilon+i0)$ also has a pole at $\varepsilon=0$.
If $R_\sigma(\varepsilon+{i}0)$ is given by Eq.\hskip2pt(\ref{EqRGap}), it follows from Eq.\hskip2pt(\ref{EqR-AM}) that ${\rm Im} \tilde{S}(\varepsilon+i0)=0$ for $|\varepsilon|<\epsilon_{\rm g}/2$ but $\varepsilon=0$.
Thus, Eq.\hskip2pt(\ref{EqKR-S}) can be described as
\begin{align}
\tilde{S}(\varepsilon+i0) &
= \frac{\tilde{\lambda}_S}{\varepsilon+i0}
+ \delta \tilde{S}(\varepsilon+i0),
\end{align}
%
where 
$\tilde{\lambda}_S = -1/\bigl[(d/d\varepsilon)R_\sigma(\varepsilon+{i}0)\bigr]>0$
and
\begin{align}\label{EqPole-SS2}
\delta \tilde{S}(\varepsilon+i0) &
=-\frac1{\pi}\left(\int_{-\infty}^{-\epsilon_{\rm g}/2}\hskip-5pt dx +\int_{+\epsilon_{\rm g}/2}^{+\infty}\hskip-5pt dx \right)\frac{\tilde{\Lambda}_S(x)}{\varepsilon+i0 -x},
\end{align}
%
where
$\tilde{\Lambda}_S(x)=\tilde{\Lambda}_S(-x)\ge 0$.
In order for a gap as large as $\epsilon_{\rm g}$ to open in the Anderson model,
$\tilde{S}(\varepsilon+{i}0)$ has to have a pole at $\varepsilon=0$ and
the residue $\tilde{\lambda}_S$ has to be so large that 
$|\varepsilon - {\rm Re}S_\sigma(\varepsilon+{i}0)|$ cannot be zero for any $\varepsilon$ of $|\varepsilon|<\epsilon_{\rm g}/2$; 
at least,\footnote{
If the multisite $\Delta\Sigma_\sigma(\varepsilon+{i}0,{\bm k})$  is continuous as a function of ${\bm k}$, Eq.\hskip2pt(\ref{EqLargeREsidue1}) or a similar one has to be satisfied for a complete gap to open in the Hubbard model, so that no complete gap can open if $W$ is infinite. If it can be discontinuous as a function of ${\bm k}$, a complete gap can open even if $W$ is infinite, as examined for a continuous model in Appendix\hskip2pt\ref{SecContinousModel}.
If a complete gap opens in a finite or infinite bandwidth model, Eq.\hskip2pt(\ref{EqLargeResidue2}) has to be satisfied in either model.
}
\begin{align}\label{EqLargeResidue2}
\tilde{\lambda}_S > \frac1{4}\left[\epsilon_{\rm g}^2 
+ 2\epsilon_{\rm g}\bigr|\delta\tilde{S}\left(\epsilon_{\rm g}/2+{i}0\right)\bigr|\right],
\end{align}
has to be satisfied.

Since $\tilde{S}(\varepsilon+{i}0)$ is the sum of $\tilde{\Sigma}_\sigma(\varepsilon+{i}0)-\mu$ and $\tilde{\Gamma}(\varepsilon+{i}0)$, as shown in Eq.\hskip2pt(\ref{EqDOS-KLT-AS}), it follows that
\begin{align}
&\tilde{\Sigma}_\sigma(\varepsilon+{i}0)-\mu 
=\frac{\tilde{\lambda}_\Sigma}{\varepsilon+i0}
+ \delta \tilde{S}_\Sigma(\varepsilon+i0), \quad
\tilde{\Gamma}(\varepsilon+{i}0) 
=\frac{\tilde{\lambda}_\Gamma}{\varepsilon+i0}
+ \delta \tilde{S}_\Gamma(\varepsilon+i0),
\end{align}
where $\tilde{\lambda}_\Sigma + \tilde{\lambda}_\Gamma=\tilde{\lambda}_S$, and that
\begin{align}
\delta \tilde{S}_\Sigma (\varepsilon+i0) &
=-\frac1{\pi}\left(\int_{-\infty}^{-\epsilon_{\rm g}/2}\hskip-5pt dx +\int_{+\epsilon_{\rm g}/2}^{+\infty}\hskip-5pt dx \right)\frac{\tilde{\Lambda}_\Sigma (x)}{\varepsilon+i0 -x},
\\
\delta \tilde{S}_\Gamma (\varepsilon+i0) &
=-\frac1{\pi}\left(\int_{-\infty}^{-\epsilon_{\rm g}/2}\hskip-5pt dx +\int_{+\epsilon_{\rm g}/2}^{+\infty}\hskip-5pt dx \right)\frac{\tilde{\Lambda}_\Gamma (x)}{\varepsilon+i0 -x},
\end{align}
where 
$\tilde{\Lambda}_\Sigma (x)+\tilde{\Lambda}_\Gamma (x)=
\tilde{\Lambda}_S (x)$ for $|x|>\epsilon_{\rm g}/2$;
in general,
$\tilde{\lambda}_\Sigma\ge 0$, $\tilde{\lambda}_\Gamma\ge 0$,  
$\tilde{\Lambda}_\Sigma (x)=\tilde{\Lambda}_\Sigma (-x)\ge 0$ for $|x|>\epsilon_{\rm g}/2$,
and
$\tilde{\Lambda}_\Gamma (x)=\tilde{\Lambda}_\Gamma (-x)=\tilde{\Delta}(x)\ge 0$ for $|x|>\epsilon_{\rm g}/2$.

An issue is whether or not $\tilde{\Gamma}(\varepsilon+{i}0)$ can have a pole at $\varepsilon=0$, i.e., whether or not $\tilde{\lambda}_\Gamma$ can be nonzero. 
There are two possible cases for $\tilde{\lambda}_\Gamma$: $\tilde{\lambda}_\Gamma =0$ and $\tilde{\lambda}_\Gamma >0$.
First, we examine the case in which $\tilde{\lambda}_\Gamma =0$, i.e., $\tilde{\lambda}_\Sigma >0$ and the single-site $\tilde{\Sigma}_\sigma(\varepsilon+{i}0)$ has a pole at $\varepsilon=0$. 
According to the analysis in Appendix\hskip2pt\ref{SecAnomaly}, $\tilde{\lambda}_\Gamma=0$ means that the multisite $\Delta\Sigma_\sigma(\varepsilon+{i}0, {\bm k})$ has no pole at $\varepsilon=0$.
The possibility of this type of solution cannot be excluded within the qualitative analysis in this section.


Second, we examine the case in which $\tilde{\lambda}_\Gamma >0$, i.e., $\tilde{\Delta}(\varepsilon+{i}0)$ includes the delta function $\delta(\varepsilon)$.
In the original definition of $\tilde{\Gamma}(\varepsilon+{i}0)$, which is given by Eq.\hskip2pt(\ref{EqHybridGamma}), we first assume that
\begin{align}\label{EqDeltaLorentz}
\tilde{\Delta}(\varepsilon) &
= \tilde{\lambda}_\Gamma \frac{\tilde{\gamma}_\Gamma}{\varepsilon^2 + \tilde{\gamma}_\Gamma^2} ,
\end{align}
where $\tilde{\gamma}_\Gamma$ is nonzero and finite such that $0< \tilde{\gamma}_\Gamma \ll \epsilon_{\rm g}$; then we eventually take the limit of $\tilde{\gamma}_\Gamma\rightarrow 0$.
It is evident that $T_{\rm K}$ is neither zero nor infinitesimal if $\tilde{\gamma}_\Gamma$ is neither zero nor infinite.
If $T_{\rm K}$ defined by Eq.\hskip2pt(\ref{EqDefTK}) is used, it follows from Eq.\hskip2pt(\ref{EqExpansionAM}) that 
\begin{align}\label{EqExpansionD}
\tilde{\Sigma}_\sigma(\varepsilon+i0) &
= \frac1{2}U
+ \left[1- \pi\tilde{\Delta}(0)/(2\tilde{W}_s k_{\rm B}T_{\rm K}) \right] \varepsilon + \cdots,
\end{align}
where $\tilde{W}_s$ is the Wilson ratio defined by Eq.\hskip2pt(\ref{EqWilsonRatio}).
This expansion is relevant for $|\varepsilon| \lesssim \min(\tilde{\gamma}_\Gamma, k_{\rm B}T_{\rm K})$.
%
It follows that
\begin{subequations}\label{EqDiscontinuity}
\begin{align}\label{EqDiscontinuity1}
\lim_{\gamma_\Gamma\rightarrow 0}
\left[\frac{d~}{d\varepsilon}{\rm Re} \tilde{\Sigma}_\sigma(\varepsilon+i0)\right]_{\varepsilon=0}&
= \lim_{\gamma_\Gamma\rightarrow 0}
\left[1- \frac{\pi\tilde{\lambda}_\Gamma}{2\tilde{W}_s k_{\rm B}T_{\rm K}\tilde{\gamma}_\Gamma}\right]
= - \infty.
\end{align}
The Kondo temperature $T_{\rm K}$ is a function of $\tilde{\gamma}_\Gamma$.
There are two possible cases for $T_{\rm K}$ in the limit of $\tilde{\gamma}_\Gamma\rightarrow 0$: $k_{\rm B}T_{\rm K} \ge \tilde{\gamma}_\Gamma$ and $k_{\rm B}T_{\rm K} < \tilde{\gamma}_\Gamma$.
If $k_{\rm B}T_{\rm K} \ge \tilde{\gamma}_\Gamma$, then
%
\begin{align}\label{EqDiscontinuity2}
\lim_{\tilde{\gamma}_\Gamma\rightarrow 0}
{\rm Re}\bigl[\tilde{\Sigma}_\sigma(-\tilde{\gamma}_\Gamma+i0)-\tilde{\Sigma}_\sigma(\tilde{\gamma}_\Gamma+i0)\bigr] =
\pi \tilde{\lambda}_\Gamma/(\tilde{W}_sk_{\rm B}T_{\rm K})>0.
\end{align}
If $k_{\rm B}T_{\rm K} < \tilde{\gamma}_\Gamma$, then
\begin{align}\label{EqDiscontinuity3}
\lim_{\tilde{\gamma}_\Gamma\rightarrow 0}
{\rm Re}\bigl[\tilde{\Sigma}_\sigma(-k_{\rm B}T_{\rm K}+i0)-\tilde{\Sigma}_\sigma(k_{\rm B}T_{\rm K}+i0)\bigr] =
\lim_{\tilde{\gamma}_\Gamma\rightarrow 0}
\pi \tilde{\lambda}_\Gamma/(\tilde{W}_s\tilde{\gamma}_\Gamma)
= + \infty.
\end{align}
\end{subequations}
According to 
Eq.\hskip2pt(\ref{EqDiscontinuity}), the real part of $\bigl[\tilde{\Sigma}_\sigma(\varepsilon+i0)\bigr]_{\tilde{\gamma}\rightarrow 0}$ is discontinuous at $\varepsilon=0$,
so that its imagery part is divergent at $\varepsilon=0$ in any possible solution of this type. 
If it continuously diverges as $\varepsilon\rightarrow 0$ such that
%
\begin{align}\label{EqDivIm1}
\lim_{\varepsilon\rightarrow 0} \bigl[{\rm Im} \tilde{\Sigma}_\sigma(\varepsilon+i0)\bigr]_{\tilde{\gamma}_\Gamma\rightarrow 0} &
= -\infty,
\end{align}
a zerogap opens, which contradicts the opening of a complete gap.
%
A complete gap opens only if it
discontinuously diverges at $\varepsilon= 0$ such that
\begin{align}\label{EqDivIm2}
\bigl[{\rm Im} \tilde{\Sigma}_\sigma(\varepsilon+i0)\bigr]_{\tilde{\gamma}_\Gamma\rightarrow 0} \propto - \delta(\varepsilon),
\end{align} 
i.e., only if $\tilde{\Sigma}_\sigma(\varepsilon+{i}0)$ has a pole at $\varepsilon=0$.
This is only possible if $T_{\rm K} \rightarrow 0\hskip1pt{\rm K}$ in the limit of $\tilde{\gamma}_\Gamma\rightarrow 0$.
Since $\tilde{\Delta}(0)\rightarrow+\infty$ as $\tilde{\gamma}_\Gamma\rightarrow 0$, i.e., $\tilde{\Delta}(0)>0$, it is probable that $T_{\rm K}=+0\hskip2pt$K rather than $T_{\rm K}=0\hskip2pt$K; at least, the entropy at $T=+0\hskip2pt$K is nonzero. 
%
This type of solution is only possible beyond S$^3$A, as studied in Appendix\hskip2pt\ref{SecAnomaly}.

In this section, we list all the possible types of solution for a complete-gapped phase 
beyond S$^3$A 
in an unspecified model.
It is shown that $\tilde{\Sigma}_\sigma(\varepsilon+{i}0)$ has to have a pole at $\varepsilon=0$ in any of them, which means that $T_{\rm K}=0\hskip2pt$K or $T_{\rm K}=+0\hskip2pt$K in it. 
Thus, 
the residual entropy or entropy at $T=+0\hskip2pt$K is nonzero in it;
the entropy at $T=+0\hskip2pt$K is $k_{\rm B}\ln 2$ per unit cell, as studied in Appendix\hskip2pt\ref{SecResidualEntropy}.


\section{Discussion} 
\label{SecDiscussion}
Since the density of states $\rho(\varepsilon)$ is independent of wave number, it is essentially a single-site or local property in any model. It is therefore reasonably anticipated that $\rho(\varepsilon)$ can be described by that of a {\it local or impurity} model even for any model, even beyond S$^3$A or DMFT; S$^3$A and DMFT are exactly equivalent to each other.
According to the Kondo-lattice theory,\cite{Mapping-1,Mapping-2,Mapping-3,toyama}
the $\rho(\varepsilon)$ of the Hubbard model can be actually described by that of the {\it impurity} Anderson model, either under or beyond S$^3$A, as shown in Eq.\hskip2pt(\ref{EqDOS2}).
Since the site-diagonal $R_{\sigma}({i}\varepsilon_l)$
includes the total $\Sigma_{\sigma}({i}\varepsilon_l,{\bm k})$, as shown in Eq.\hskip2pt(\ref{EqGreenR}), the mapping condition of Eq.\hskip2pt(\ref{EqMap4}) depends on
the multisite $\Delta\Sigma_{\sigma}({i}\varepsilon_l,{\bm k})$, i.e., the hybridization energy $\tilde{\Delta}(\varepsilon)$ of the Anderson model to be mapped depends on $\Delta\Sigma_{\sigma}({i}\varepsilon_l,{\bm k})$;
the multisite $\Delta\Sigma_\sigma({i}\varepsilon_l,{\bm k})$ is approximately or rigorously considered beyond S$^3$A, although it is ignored under S$^3$A. Beyond S$^3$A,
therefore, all of the $\tilde{\Delta}(\varepsilon)$, $\tilde{\Sigma}_\sigma({i}\varepsilon_l)$, and $\Delta\Sigma_{\sigma}({i}\varepsilon_l,{\bm k})$ have to be self-consistently calculated with each other to satisfy the mapping condition of Eq.\hskip2pt(\ref{EqMap4}). 
The effects of intersite correlations, which are described in terms of the multisite $\Delta\Sigma_\sigma({i}\varepsilon_l, {\bm k})$ of the Hubbard model, can be described in terms of the self-consistent $\tilde{\Delta}(\varepsilon)$ of the Anderson model, at least, as regards any local property such as $\rho(\varepsilon)$.

The residual entropy or entropy at $T=+0\hskip2pt$K is nonzero for any possible complete-gapped phase either under or beyond S$^3$A or DMFT;
in particular, the entropy at $T=+0\hskip2pt$K is $k_{\rm B}\ln 2$ per unit cell, as studied in Appendix\hskip2pt\ref{SecResidualEntropy}.
In Brinkman and Rice's theory,\cite{brinkman} the metal-insulator transition appears continuous.
According to the analysis in the present paper, it is discontinuous such that the residual entropy or entropy at $T=+0\hskip2pt$K jumps at the transition.

Theories based on DMFT\cite{kotliar,moeller,bulla} are consistent with Brinkman and Rice's theory except for the hysteresis in DMFT.
Since the transition is discontinuous, it is reasonable that the hysteresis appears.
Under DMFT, a metal-insulator transition occurs at $U_{c_2}$ as $U$ increases, while it occurs at $U_{c_1}$, which is smaller than $U_{c_2}$, as $U$ decreases.
According to the analysis in the present paper, the hysteresis can be explained in the following way: the Kondo temperature $T_{\rm K}$ continuously vanishes at $U_{c_2}$ as $U$ increases, while it discontinuously becomes nonzero at $U_{c_1}$ as $U$ decreases; 
and the residual entropy or entropy at $T=+0\hskip2pt$K discontinuously becomes nonzero at $U_{c_2}$ as $U$ increases, while it discontinuously becomes zero at $U_{c_1}$ as $U$ decreases.
The jump of $T_{\rm K}$ at $U_{c_1}$ means that the internal energy also jumps at $U_{c_1}$.

In general, an infinitely degenerate ground state is unstable even in the presence of an infinitesimal perturbation, except for a symmetry-broken state in which a continuous symmetry is broken and rigidity appears.\cite{AndersonText} 
This is why the third law of thermodynamics is valid.
It is therefore doubtful that a gapped phase with no symmetry broken, in which the residual entropy or entropy at $T=+0\hskip2pt$K is nonzero, is stable under or beyond S$^3$A in the presence of an infinitesimal perturbation due to the communication of electrons between the Hubbard model and an electron reservoir, i.e., in the grand canonical ensemble, as reported in a previous paper.\cite{toyama}

In two dimensions and higher, symmetry can be broken when the temperature $T$ is low enough and, at least, at the absolute zero Kelvin.
An infinitely degenerate ground state in which the third law of thermodynamics is broken is unstable against an ordered state such as an antiferromagnetic state.
The third law of thermodynamics is valid in any ordered ground state because rigidity appears in it.\cite{AndersonText} 

The Kondo temperature $T_{\rm K}$ or $k_{\rm B}T_{\rm K}$ is the energy scale of single-site quantum spin  fluctuations.
%
Electrons behave like local moments in a high-$T$ phase at $T\gg T_{\rm K}$, because single-site thermal spin fluctuations overcome single-site quantum ones.
The high-$T$ phase at $T\gg T_{\rm K}$ is in a sense a paramagnetic type of the Mott insulator, which is not the ground state.
If an antiferromagnetic gap opens, the ground state can also be an insulator \cite{slater}.
A local-moment type of antiferromagnetic insulator, in which the N\'{e}el temperature $T_{\rm N}$ is as high as $T_{\rm N} \gg T_{\rm K}$, is an antiferromagnetic type of the Mott insulator, because electrons behave like local moments in its paramagnetic phase at $T> T_{\rm N}$.
If electrons are not half-filled and the nesting of the Fermi surface is not sharp, $T_{\rm N}$ is low or it vanishes.
If $T_{\rm N}$ is nonzero and $T_{\rm N}\ll T_{\rm K}$, electrons are itinerant in a paramagnetic phase at $T_{\rm N}<T\ll T_{\rm K}$, so that the magnetism at $T\le T_{\rm N}$ is itinerant-electron magnetism.
If $T_{\rm K}$ is high enough, it is possible that the ground state is superconducting.\cite{FJO-SC2}

The Hubbard model in one dimension is particular, because no symmetry can be broken in it,\cite{mermin} and 
because the Bethe-ansatz solution was given by Lieb and Wu.\cite{lieb-wu}
We denote the ground-state energy in the canonical ensemble by $E_{\rm G}(N)$ as a function of the number $N$ of electrons.
Two {\it chemical potentials} are defined for the ground state of $N$ electrons in the canonical ensemble: 
$\mu_{+}(N)= E_{\rm G}(N+1)- E_{\rm G}(N)$
for the addition of an electron and 
$\mu_{-}(N)= E_{\rm G}(N)- E_{\rm G}(N-1)$
for the removal of an electron. 
A {\em gap} is defined by 
$\epsilon_{\rm g} (N) =\mu_{+}(N)-\mu_{-}(N)$.
%
According to the Bethe-ansatz solution,
$\epsilon_{\rm g} (N)=0$ for $N\ne L$, while
$\epsilon_{\rm g} (N)>0$ for $N= L$ 
for any nonzero $U$; i.e.,
a {\it gap} opens in the spectrum of adding or removing the whole of a single electron in the exactly half-filled case.
On the other hand, the third law of thermodynamics is not broken in the ground state of the Bethe-ansatz solution.
If the opening of a gap in the spectrum of adding or removing an electron in the canonical ensemble is equivalent to that in the single-particle excitation spectrum in the grand canonical ensemble, the conclusion of the present paper and the argument based on the Bethe-ansatz solution are contradictory to each other.


A possible explanation for the contradiction is that the Kondo-lattice theory, which is none other than the $1/D$ expansion theory from infinite dimensions, is inapplicable to the opposite limit of one dimension. 
Since the Kondo-lattice theory is rigorous in infinite dimensions as S$^3$A is, it is applicable at least to $D\rightarrow +\infty$ dimensions.\footnote{
If the multisite $\Delta\Sigma_\sigma(\varepsilon+i0,{\bm k})$ as well as the single-site $\tilde{\Sigma}_\sigma(\varepsilon+i0)$ is rigorously calculated, the Kondo-lattice theory that includes no conventional Weiss mean field, if it is applicable and relevant, is rigorous for any $D$ dimensions, but within the constraint Hilbert subspace in which no symmetry is allowed to be broken the same as $S^3$A for infinite dimensions.
} 
If the explanation is right to the point, therefore, there is a critical $D_c$ such that the Kondo-lattice theory is applicable to $D\ge D_c$ dimensions while it is inapplicable to $1\le D< D_c$ dimensions.
It is desirable to determine how large or small the critical $D_c$ is. It is also desirable to clarify why the Kondo-lattice theory is inapplicable to $1\le D< D_c$ dimensions and, in particular, why the mapping of the single-site self-energy of the Hubbard model to the local self-energy of the Anderson model is impossible for $1\le D< D_c$.
Even if the Kondo-lattice theory is really inapplicable to $1\le D< D_c$ dimensions, 
a necessary condition for the opening of a gap in the single-particle excitation spectrum is that
the total self-energy $\Sigma_\sigma(\varepsilon+{i}0,{\bm k})$ has a pole at $\varepsilon=0$ at least for ${\bm k}={\bm k}_{\rm F}$ on the hypersurface defined by Eq.\hskip2pt(\ref{EqHyper-surface}), as discussed in the last paragraph in \S\ref{SecKL-theory}.
Since the self-energy $\Sigma_\sigma(\varepsilon+{i}0,{\bm k})$ having a pole at $\varepsilon=0$ implies or at least suggests that there are degenerate zero-energy bosonic excitations, which are responsible for the divergent imaginary part of the self-energy,
it is desirable to elucidate why and how the residual entropy can be zero even in such an anomalous situation.

Another possible explanation for the contradiction is that the opening of a gap in the canonical ensemble is not equivalent to that in the single-particle excitation spectrum in the grand canonical ensemble.
%
Although the above well-known argument by Lieb and Wu is certainly based on the exact result of $\left[\epsilon_{\rm g}(N)\right]_{N=L}>0$,
which is obtained by the Bethe-ansatz solution, the inequality of $\epsilon_{\rm g}(N)>0$ is not a sufficient condition for the ground state being an insulator;
for example, the very same inequality is satisfied in a metallic fine particle if the long-range Coulomb interaction is explicitly considered.
Thus, the possibility that the half-filled ground state is a metal is not excluded by the Bethe-ansatz solution itself.
It should be examined whether a gap opens in the single-particle excitation spectrum in the grand canonical ensemble.
If the Kondo-lattice theory is applicable to one dimension, a combination of the ground-state degeneracy of the Bethe-ansatz solution and the conclusion of the present paper means that no complete gap opens in the single-particle excitation spectrum.
It is desirable to study the conductivity itself in the presence of electrodes, which plays a role of an electron reservoir for the Hubbard model. 

In the continuous limit of $a\rightarrow 0$, where $a$ is the lattice constant,
the Hubbard model in one dimension is substituted for the Tomonaga-Luttinger model with umklapp terms;\cite{Tomonaga,Luttinger,Mattis} 
and the Tomonaga-Luttinger model is further substituted for a boson model, in which a charge gap opens.\cite{Solym} 
Since the Tomonaga-Luttinger model is a continuous or nonlattice   model, the Kondo-lattice theory is inapplicable to it, i.e., the Kondo effect is irrelevant in it, as discussed in Appendix\hskip2pt\ref{SecContinousModel}.
If the Kondo-lattice theory is applicable to the Hubbard model in one dimension, 
the opening of a complete gap in the Tomonaga-Luttinger or Boson model, in which the Kondo effect is irrelevant, is not a proof that contradicts the conclusion of the present paper for the Hubbard model, in which the Kondo effect is relevant.


\section{Conclusion} 
\label{SecConclusion}
%
The self-energy of the Hubbard model is decomposed into the single-site and multisite ones.
According to the Kondo-lattice theory,
the single-site self-energy is equal to that of the mapped Anderson model that is self-consistently determined in a way such that
the density of states of the Hubbard model per unit cell is equal to  that of the Anderson model.
If the multisite self-energy is ignored under the supreme single-site approximation (S$^3$A) or if it is approximately or rigorously considered beyond S$^3$A,
the density of states per unit cell is simply given by
%
\begin{align}
\rho(\varepsilon) = - \frac1{\pi} {\rm Im}
\frac1{\varepsilon+i0 - \tilde{S}(\varepsilon+{i}0)},
\end{align}
where 
%
$\tilde{S}(\varepsilon+{i}0)= 
\tilde{\epsilon}_d -\tilde{\mu}+ \tilde{\Sigma}_\sigma(\varepsilon+{i}0)+\tilde{\Gamma}(\varepsilon+{i}0)$,  
%
where $\tilde{\epsilon}_d$, $\tilde{\mu}$, $\tilde{\Sigma}_\sigma(\varepsilon+{i}0)$, and $\tilde{\Gamma}(\varepsilon+{i}0)$ are the level of localized electrons, the chemical potential, the self-energy, and the hybridization term, respectively, of the mapped Anderson model; and $\tilde{\mu}-\tilde{\epsilon}_d=\mu-\epsilon_d$, where $\mu$ and $\epsilon_d$ are the chemical potential and band center, respectively, of the Hubbard model.
The Kondo temperature $T_{\rm K}$ or $k_{\rm B}T_{\rm K}$ of the mapped Anderson model is also the low-energy scale of single-site quantum spin fluctuations in  the Hubbard model. 


In the exactly half-filled case of the Hubbard model on a bipartite lattice, ${\rm Re}\tilde{S}(+{i}0)=0$
because of the particle-hole symmetry.
A complete gap as large as $\epsilon_{\rm g}$ can open in $\rho(\varepsilon)$,
if and only if $\tilde{S}(\varepsilon+{i}0)$ has a pole at $\varepsilon=0$, the residue of the pole is so large that 
$|\varepsilon - {\rm Re}S_\sigma(\varepsilon+{i}0)|$ cannot be zero for any $|\varepsilon|<\epsilon_{\rm g}/2$, 
and ${\rm Im}\tilde{S}(\varepsilon+{i}0)=0$ for any $|\varepsilon|<\epsilon_{\rm g}/2$ but $\varepsilon= 0$.
Such an anomalous $\tilde{S}(\varepsilon+{i}0)$ is possible if and only if $T_{\rm K}=0\hskip2pt$K or $T_{\rm K}=+0\hskip2pt$K.
If $T_{\rm K}=0\hskip2pt$K, the residual entropy is nonzero. 
If $T_{\rm K}=+0\hskip2pt$K, the ground state is a singlet, while the entropy at $T= +0\hskip2pt$K is nonzero; the entropy discontinuously vanishes at $T=0\hskip2pt$K with decreasing $T$. Thus, the Mott insulator with no symmetry broken, if it is possible, is characterized by nonzero residual entropy or nonzero entropy at $T=+0\hskip2pt$K.


%

The conclusion of the present paper is consistent with Brinkman and Rice's theory, in which the specific-heat coefficient diverges as the onsite $U$ approaches a transition point from a metallic phase and no symmetry is broken in an insulating phase; a combination of these two properties means that the residual entropy or entropy at $T=+0\hskip2pt$K is nonzero in the insulating phase. 
It is also consistent with the appearance of hysteresis in the dynamical mean-field theory.

According to the well-known argument based on the Bethe-ansatz solution, on the other hand, the half-filled ground state
in one dimension is the Mott insulator for any nonzero onsite $U$, although the third law of thermodynamics is not broken in it; thus,
the well-known argument and the conclusion of the present paper contradict each other.
A possible explanation for the contradiction 
is that the Kondo-lattice theory, which is none other than the $1/D$ expansion theory from infinite dimensions, is inapplicable to one dimension.
The other possible explanation is that 
a gap in the spectrum of adding or removing an electron in the canonical ensemble, which is given by the Bethe-ansatz solution, is different from a gap in the single-particle excitation spectrum in the grand canonical ensemble, which is studied in the present paper.
Since the opening of the gap in the canonical ensemble is not a sufficient condition for the ground state being an insulator,  
the well-known argument is not a proof that the half-filled ground state in one dimension is the Mott insulator.
If the Kondo-lattice theory is applicable to one dimension, a combination of the conclusion of the present paper and the ground-state degeneracy of the Bethe-ansatz solution means that the half-filled ground state in one dimension is not the Mott insulator.
It is desirable to elucidate which explanation is true in the near future.

\appendix 
\section{Anomalies of Gapped Phases}
\label{SecAnomaly}
In order for a complete gap as large as $\epsilon_{\rm g}$ to open, the single-site $\tilde{\Sigma}_\sigma(\varepsilon+{i}0)$ has to have a pole at $\varepsilon=0$, as examined in \S\S\hskip1pt\ref{SecFiniteW} and \hskip1pt\ref{SecKLT}, so that
the total $\Sigma_\sigma(\varepsilon+i0,{\bm k})$ also has to have a pole at $\varepsilon=0$. 
However, it does not have to have two or more poles in the gap region of $|\varepsilon|<\epsilon_{\rm g}$/2, as discussed below.

If the total $\Sigma_\sigma(\varepsilon+i0,{\bm k})$ has two or more poles in the region of $|\varepsilon|<\epsilon_{\rm g}/2$,
there is at least a pair of adjacent poles: e.g., one at $\varepsilon=p_1$ and the other at $\varepsilon=p_2$, where $-\epsilon_{\rm g}/2<p_1<p_2<\epsilon_{\rm g}/2$. Then, it follows that
\begin{align}
\lim_{\varepsilon\rightarrow p_1+0} {\rm Re} \Sigma_\sigma(\varepsilon+i0,{\bm k}) = + \infty, \quad
\lim_{\varepsilon\rightarrow p_2-0} {\rm Re} \Sigma_\sigma(\varepsilon+i0,{\bm k}) = - \infty.
\end{align}
Since $\Sigma_\sigma(\varepsilon+i0,{\bm k})$ has no singularity between poles and is continuous between poles, 
there is a zero point $z_\varepsilon$ between poles such that it satisfies
\begin{align}
{\rm Re}\bigl[z_\varepsilon +\mu -E({\bm k})- \Sigma_\sigma(z_\varepsilon+i0,{\bm k})\bigr]=0.
\end{align}
Since the zero point $z_\varepsilon$ is in the gap region of $|\varepsilon|<\epsilon_{\rm g}/2$, it follows that
\begin{align}
{\rm Im}\bigl[z_\varepsilon +\mu - E({\bm k})-\Sigma_\sigma(z_\varepsilon+i0,{\bm k})\bigr]=0.
\end{align}
Since the density of states $\rho(\varepsilon)$ of the Hubbard model is given by Eq.(\ref{EqDOS}) with Eq.(\ref{EqGreenR}),
an in-gap state inevitably appears at each zero point $\varepsilon=z_\varepsilon$ between each pair of adjacent poles.
This conclusion contradicts the opening of a complete gap.
Thus, the total self-energy $\Sigma_\sigma(\varepsilon+i0,{\bm k})$ does not have to have two or more poles in the gap region of $|\varepsilon|<\epsilon_{\rm g}/2$ for a complete gap as large as $\epsilon_{\rm g}$ to open.

A complete gap as large as $\epsilon_{\rm g}$ opens in the Hubbard model if and only if single-site and multisite self-energies are given by 
\begin{subequations}\label{EqPole-S0}
\begin{align}
&\hskip15pt 
\tilde{\Sigma}_\sigma({i}\varepsilon_l)-\mu 
=\frac{\tilde{\lambda}_\Sigma}{{i}\varepsilon_l}
-\frac1{\pi}\left(\int_{-\infty}^{-\epsilon_{\rm g}/2}\hskip-5pt dx +\int_{+\epsilon_{\rm g}/2}^{+\infty}\hskip-5pt dx \right)\frac{\tilde{\Lambda}_\Sigma (x)}{{i}\varepsilon_l -x},
\\ &
\Delta\Sigma_\sigma({i}\varepsilon_l,{\bm k}) 
=\frac{\lambda_{\Delta\Sigma}({\bm k})}{{i}\varepsilon_l}
-\frac1{\pi}\left(\int_{-\infty}^{-\epsilon_{\rm g}/2}\hskip-5pt dx +\int_{+\epsilon_{\rm g}/2}^{+\infty}\hskip-5pt dx \right)\frac{\Lambda_{\Delta\Sigma} (x;{\bm k})}{{i}\varepsilon_l -x},
\end{align}
\end{subequations}
where $\tilde{\lambda}_\Sigma + \tilde{\lambda}_{\Delta\Sigma}({\bm k})$ is nonzero and large enough; $\tilde{\lambda}_\Sigma \ge 0$ and $\tilde{\lambda}_{\Delta\Sigma}({\bm k}) \ge 0$, in general.
According to Eqs.\hskip2pt(\ref{EqGreenR}) and (\ref{EqPole-S0}), 
\begin{align}
R_\sigma({i}\varepsilon_l)  = 
{i}\varepsilon_l \left[\frac1{L}\sum_{\bm k} \frac1{\tilde{\lambda}_\Sigma+ \lambda_{\Delta\Sigma}({\bm k}) } \right] + O\left[\left({i}\varepsilon_l\right)^3\right],
\end{align}
in the limit of $|\varepsilon_l|\rightarrow 0$. According to the mapping condition of Eq.\hskip2pt(\ref{EqMap4}), 
\begin{align}
\tilde{\Delta}(\varepsilon) =
(\tilde{\lambda}_\Gamma/\pi) \delta(\varepsilon),
\end{align}
for $|\varepsilon|<\epsilon_{\rm g}/2$, where  
\begin{align}
\tilde{\lambda}_\Gamma =
\left\{\begin{array}{cc}
\displaystyle
\left[\frac1{L}\sum_{\bm k} \left(
\frac1{\tilde{\lambda}_\Sigma+ \lambda_{\Delta\Sigma}({\bm k})} - \frac1{ \tilde{\lambda}_\Sigma}
\right)\right]^{-1}, & \tilde{\lambda}_\Sigma>0, \hskip3pt \lambda_{\Delta\Sigma}({\bm k})\ge 0, \\
\displaystyle
\left[\frac1{L}\sum_{\bm k} 
\frac1{\lambda_{\Delta\Sigma}({\bm k})} \right]^{-1}, & 
\tilde{\lambda}_\Sigma=0, \hskip3pt \lambda_{\Delta\Sigma}({\bm k})>0. 
\end{array} \right. 
\end{align}
If $\lambda_{\Delta\Sigma}({\bm k})=0$ or $\Delta\Sigma_\sigma(\varepsilon+{i}0,{\bm k})$ has no pole at $\varepsilon=0$, then $\tilde{\lambda}_\Gamma=0$, i.e., $\tilde{\Gamma}(\varepsilon+{i}0)$ has no pole at $\varepsilon=0$.
If $\lambda_{\Delta\Sigma}({\bm k})>0$ or $\Delta\Sigma_\sigma(\varepsilon+{i}0,{\bm k})$  
has a pole at $\varepsilon=0$,  then $\tilde{\lambda}_\Gamma>0$, i.e., $\tilde{\Gamma}(\varepsilon+{i}0)$ has a pole at $\varepsilon=0$.
According to the analysis in \S\ref{SecKLT}, if $\tilde{\Gamma}(\varepsilon+{i}0)$ has a pole at $\varepsilon=0$, 
the single-site $\tilde{\Sigma}_\sigma(\varepsilon+{i}0)$ also has a pole at $\varepsilon=0$.
Thus, any self-consistent solution characterized by 
$\tilde{\lambda}_\Sigma=0$ and $\lambda_{\Delta\Sigma}({\bm k})>0$ is impossible.

Next, we assume that the single-site $\tilde{\Sigma}_\sigma(\varepsilon+{i}0)$ is divergent at $\varepsilon=0$ such that
\begin{subequations}
\begin{align}
\lim_{\varepsilon\rightarrow 0}
\bigl|\tilde{\Sigma}_\sigma(\varepsilon+{i}0)\bigr|=+\infty,
\quad 
\bigl|\tilde{\Sigma}_\sigma(+{i}0)\bigr|=+\infty,
\end{align}
while the multisite $\Delta\Sigma_\sigma(\varepsilon+{i}0,{\bm k})$ is finite and continuous at $\varepsilon=0$ and
\begin{align}\label{EqDeltaS-Vanish}
{\rm Im} \bigl[\Delta\Sigma_\sigma(\varepsilon+{i}0,{\bm k})\bigr]=0,
\end{align}
\end{subequations}
for any $|\varepsilon|<\epsilon_{\rm g}/2$ and any ${\bm k}$.
We define 
\begin{align}
&\tilde{g}(\varepsilon) = \frac1{\varepsilon +{i}0 +\mu -\tilde{\Sigma}_\sigma(\varepsilon+{i}0)}, \quad
Q_n(\varepsilon) = \frac1{L}\sum_{\bm k} \bigl[E({\bm k})+\Delta\Sigma_\sigma(\varepsilon+{i}0,{\bm k})\bigr]^n.
\end{align}
It is easy to see that
\begin{align}\label{EqVanishG}
\lim_{\varepsilon\rightarrow 0}\bigl|\tilde{g}(\varepsilon)\bigr|=0,
\quad
\bigl|\tilde{g}(0)\bigr|=0,
\end{align}
$Q_n(\varepsilon)$ is real for $|\varepsilon|<\epsilon_{\rm g}/2$, and $Q_n(\varepsilon)=(-1)^n Q_n(-\varepsilon)$ for $|\varepsilon|<\epsilon_{\rm g}/2$.
According to Eq.\hskip2pt(\ref{EqGreenR}), 
\begin{align}
\frac1{R_\sigma(\varepsilon+{i}0)} &=
\frac1{\tilde{g}(\varepsilon)} 
- Q_1(\varepsilon)
- \tilde{g}(\varepsilon) \left[Q_2(\varepsilon)-Q_1^2(\varepsilon)\right]
+O\left[ \tilde{g}^2(\varepsilon) \right],
\end{align} 
for a sufficiently small $|\varepsilon|$.
According to the mapping condition of Eq.\hskip2pt(\ref{EqMap4}), 
\begin{align}\label{EqDeltaApp}
\tilde{\Delta}(\varepsilon) =
- {\rm Im}\bigl\{\tilde{g}(\varepsilon) \bigl[Q_2(\varepsilon)-Q_1^2(\varepsilon)\bigr]
+O \bigl[\tilde{g}^2(\varepsilon) \bigr]
\bigr\},
\end{align}
for a sufficiently small $|\varepsilon|$.
It follows from Eqs.\hskip2pt(\ref{EqVanishG}) and (\ref{EqDeltaApp}) 
 that 
\begin{align}\label{EqVanishDelta}
\lim_{\varepsilon\rightarrow 0}\tilde{\Delta}(\varepsilon)=0,
\quad \tilde{\Delta}(0)=0.
\end{align}
In particular, the multisite $\Delta\Sigma_\sigma(\varepsilon+{i}0,{\bm k})$ vanishes under S$^3$A. Therefore, Eq.\hskip2pt(\ref{EqVanishDelta}) has to be satisfied in any possible complete-gapped or zerogapped phase under S$^3$A.

\section{Entropy at $T=+0\hskip2pt$K of Possible Complete-Gapped Phases}
\label{SecResidualEntropy}
%
If no symmetry is broken, the thermodynamic potential is given by\cite{LW1,LW2}
\begin{align}\label{EqLWOmegaH}
\Omega &=
-k_{\rm B}T \sum_{l}\sum_{{\bm k}\sigma} e^{{i}\varepsilon_l 0^+}
\Bigl\{\ln \bigl[-1/G_\sigma({i}\varepsilon_l,{\bm k}) \bigr]
%
+ G_\sigma ({i}\varepsilon_l,{\bm k}) \Sigma_\sigma({i}\varepsilon_l,{\bm k})\Bigr\}
+ \Omega^\prime,
\end{align}
where $\Omega^\prime$ is the contribution of all the closed connected Feynman diagrams but with each electron line replaced by its renormalized one and is given by
\begin{align}
\Omega^\prime &=
k_{\rm B}T \sum_{\nu} \sum_{l}\sum_{k\sigma} \frac1{2\nu}
G_\sigma({i}\varepsilon_l, {\bm k})
\Sigma_{\nu\sigma}({i}\varepsilon_l, {\bm k}),
\end{align}
where $\Sigma_{\nu\sigma}({i}\varepsilon_l, k)$ is the $\nu$th order self-energy in $U$, where only the $U$ occurring in explicit interactions of the skeleton diagram is used to determine the order.
Since the thermodynamic potential $\Omega$ is stationary with respect to the self-energy such that
$\partial \Omega/\partial \Sigma_\sigma({i}\varepsilon_l,{\bm k}) =0$ in the functional derivative, the temperature dependence of $\Sigma_\sigma({i}\varepsilon_l,{\bm k})$ does not have to be considered in calculating entropy $S=-\partial \Omega/\partial T$ except for that coming from ${i}\varepsilon_l$ or the $l$ sum.

As discussed by Luttinger and Ward,\cite{LW1,LW2}
the first correction to the $T=0$ value of $\Omega$ comes only from the difference between the $l$ sum and what we would obtain if we replaced them by integrals in each electron line.
Then, the first correction to the $T=0$ value of $\Omega^\prime$ is equal to that of
\begin{align}\label{EqOmega1}
\bar{\Omega}^{\prime}&
= k_{\rm B}T  
\sum_{\nu} \sum_{l}\sum_{k\sigma}G_\sigma({i}\varepsilon_l, {\bm k})
\Sigma_{\nu\sigma}({i}\varepsilon_l, {\bm k})%
\nonumber \\ &
= k_{\rm B}T \sum_{l} \sum_{k\sigma} 
G_\sigma({i}\varepsilon_l, {\bm k})
\Sigma_{\sigma}({i}\varepsilon_l, {\bm k}),
\end{align}
so that the first correction to the $T=0$ value of the total $\Omega$ is equal to that of
\begin{align}\label{EqOmega2}
\bar{\Omega}(T) &= 
-k_{\rm B}T \sum_{l}\sum_{{\bm k}\sigma} e^{{i}\varepsilon_l 0^+}
\ln \bigl[ -1/G_\sigma({i}\varepsilon_l,{\bm k})\bigr]
\nonumber \\ &=
\frac1{\pi} \sum_{{\bm k}\sigma}
\int_{-\infty}^{+\infty} d\epsilon f(\epsilon)
{\rm Im} \ln \bigl[ 
-\varepsilon-{i}0 + E({\bm k}) + \Sigma_\sigma(\varepsilon+{i}0,{\bm k}) - \mu
\bigr],
\end{align}
%
where $f(\epsilon)=1/\bigl[e^{\epsilon/(k_{\rm B}T)}+1\bigr]$; 
in Eqs.\hskip2pt(\ref{EqOmega1}) and (\ref{EqOmega2}),
$T=0\hskip2pt$K is assumed in the self-energy except for the $l$ sum.

For any possible complete-gap phase either under or beyond S$^3$A,
it follows according to Eq.\hskip2pt(\ref{EqPole-S0}) that
\begin{align}
\Sigma_\sigma(\varepsilon+{i}0,{\bm k}) - \mu
=\frac{\lambda_{\Sigma}({\bm k})}{\varepsilon+{i}0}
-\frac1{\pi}\left(\int_{-\infty}^{-\epsilon_{\rm g}/2}\hskip-5pt dx +\int_{+\epsilon_{\rm g}/2}^{+\infty}\hskip-5pt dx \right)\frac{\Lambda_{\Sigma} (x;{\bm k})}{\varepsilon+{i}0 -x},
\end{align}
where $\lambda_{\Sigma}({\bm k})=\tilde{\lambda}_\Sigma+\Delta\lambda_{\Delta\Sigma}({\bm k})>0$ and $\Lambda_{\Sigma} (x;{\bm k})=\tilde{\Lambda}_\Sigma+\Delta\Lambda_{\Delta\Sigma} (x;{\bm k})>0$.
Then, it follows that
\begin{align}
{\rm Im} \ln \bigl[ 
-\varepsilon-{i}0 + E({\bm k}) + \Sigma_\sigma(\varepsilon+{i}0,{\bm k}) - \mu
\bigr] &=
\left\{\begin{array}{cc}
0, & -\epsilon_{\rm g}/2< \varepsilon<0\vspace{2pt},\\
-\pi/2, & \varepsilon=0 \vspace{2pt},\\
-\pi, & 0< \varepsilon<+\epsilon_{\rm g}/2,
\end{array} \right. 
\end{align}
for $|\varepsilon|<\epsilon_{\rm g}/2$.
The entropy at $T=+0\hskip2pt$K per unit cell is as large as
\begin{align}
S/L &= 
-\lim_{T\rightarrow 0\hskip1pt{\rm K}}\frac1{L}\frac{\partial \bar{\Omega}(T)}{\partial T}
=
\lim_{T\rightarrow 0\hskip1pt{\rm K}} \frac{\partial}{\partial T}
\int_{0}^{+\epsilon_{\rm g}/2} d\epsilon f(\epsilon)
= k_{\rm B}\ln 2,
\end{align}
for any possible complete-gap phase either under or beyond S$^3$A. 

\section{Continuous Model in One Dimension}
\label{SecContinousModel}
We consider a finite or infinite bandwidth model in one dimension in the continuous limit of
$a\rightarrow 0$, with any of $k_{\rm F} a = \pi/2$,
$\hbar v_{\rm F} = \bigl[(\partial/\partial k)E(k)\bigr]_{k=k_{\rm F}}$, $g= U a$, and $L a$ being kept constant; $v_{\rm F}$ is the Fermi velocity and $g$ is the coupling constant.
We assume the thermodynamic limit of $L a\rightarrow +\infty$.


If $a$ is nonzero and finite, the single-site Green function is given by
\begin{align}
R_\sigma(\varepsilon+i0) &
= 
\frac{a}{2\pi} \int_{-\pi/a}^{+\pi/a} dk \frac1
{\varepsilon+i0 - E(k) - S(\varepsilon+i0,k)},
\end{align}
where 
\begin{align}
S(\varepsilon+i0,k)=\Sigma_\sigma(\varepsilon+i0,k)-\mu.
\end{align}
Since it vanishes in the continuous limit of $a\rightarrow0$, the single-site self-energy is never well-defined in any continuous model. Thus, the Kondo-lattice theory is inapplicable to any continuous model, so that the Kondo effect is irrelevant in it. 

When the infinite bandwidth model defined in \S\ref{SecModel} is considered, for example,
the bare density of states per unit length is given by\footnote{
If $\bar{\rho}_0(E)=e^{-\alpha |E|/(\hbar v_{\rm F})}/(\pi \hbar v_{\rm F})$ with $\alpha=+0$ is assumed instead of Eq.\hskip2pt(\ref{EqRhoBar}), it is none other than the band cutoff introduced into the boson model.} 
\begin{align}\label{EqRhoBar}
\bar{\rho}_0(E) = 
\frac1{a} \rho_0(E)
= \frac1{\pi a} \frac{|t_\infty|}{E^2 +|t_\infty| ^2}
= \frac1{\pi \hbar v_{\rm F}} \frac1{1+ [E a/(\hbar v_{\rm F})]^2},
\end{align}
where $a=+0$.
Since $E(k)$ is a continuous function of $k$, $k$ is also a continuous function of $E$. 
Thus, $S[\varepsilon+i0,k(E)]$ is a function of $E$; it is simply denoted by $S(\varepsilon+i0,E)$.
The density of state per unit length is given by
\begin{align}\label{EqDOS-Continuous}
\bar{\rho}(\varepsilon) &
= \frac1{\pi} \int_{-\infty}^{+\infty} dE \bar{\rho}_0(E) \frac1{
\varepsilon+i0 - E - S(\varepsilon+i0,E)}.
\end{align}
Because of the particle-hole symmetry,
%
${\rm Re} S(+i0,0) = 0$.
%
It is easy to see that unless ${\rm Im}S(\varepsilon+i0,0)$ is divergent at $\varepsilon=0$, then $\bar{\rho}(0)>0$ and not $\bar{\rho}(0)=0$.

We examine a possible complete-gapped phase whose gap is as large as $\epsilon_{\rm g}$: $\bar{\rho}(\varepsilon)=0$ for $|\varepsilon|<\epsilon_{\rm g}/2$.
$S(\varepsilon+i0,E)$ as a function of $E$ may be continuous or discontinuous.
First, we assume that $S(\varepsilon+i0,E)$ is continuous as a function of $E$.
According to an analysis similar to that in Appendix\hskip2pt\ref{SecAnomaly}, 
$S(\varepsilon+i0,E)$ does not have to have two or more poles in the region of $|\varepsilon|<\epsilon_{\rm g}/2$ for a complete gap to open.
Thus, it has to have zero or only one pole in the region, so that
\begin{align}
S(\varepsilon+i0,E)&
= \frac{\lambda(E)}{\varepsilon+i0 - p(E)}
- \frac1{\pi} \left(
\int_{-\infty}^{-\epsilon_{\rm g}/2} \hskip-3pt dx 
+ \int_{+\epsilon_{\rm g}/2}^{+\infty} \hskip-3pt dx\right)
\frac{\Lambda(x;E)}{\varepsilon+i0-x},
\end{align}
where $\lambda(E) = \lambda(-E)\ge 0$, 
$p(E) = - p(-E)$,  and 
$\Lambda(x;E) = \Lambda(-x;-E)>0$;
in particular,
%
$\lambda(0)>0$ and $p(0)=0$.
%
If the pole $p(E)$ depends on $E$, $S(\varepsilon+i0,E)$ as a function of $E$ is discontinuous at $E$ such that it satisfies $p(E)=\varepsilon$. Thus, $p(E)=0$ because of the constraint imposed by the assumption above.
It is evident that
\begin{align}
\lim_{E\rightarrow +\infty} \left[E + S(\varepsilon+i0,E) \right]= + \infty, \quad
\lim_{E\rightarrow -\infty} \left[E + S(\varepsilon+i0,E)\right]= -\infty.
\end{align}
Then, there is a zero point $z_E$ such that it satisfies
\begin{align}\label{EqSolution1}
{\rm Re}\left[\varepsilon - z_E - S(\varepsilon+i0,z_E) \right]=0,
\end{align}
for any finite $\varepsilon$ but $\varepsilon=0$.
The zero point $z_E$ is a function of $\varepsilon$; the function is denoted by $z_E(\varepsilon)$.
If $|\varepsilon|<\epsilon_{\rm g}/2$ but $\varepsilon\ne 0$, 
\begin{align}\label{EqSolution2}
{\rm Im}\left\{\varepsilon- z_E(\varepsilon) - S\left[\varepsilon+i0,z_E(\varepsilon)\right] \right\}=0.
\end{align}
Then, $\bar{\rho}(\varepsilon)>0$ for any $\varepsilon$ but $\varepsilon=0$. This conclusion contradicts the opening of a complete gap. 
If the bandwidth is infinite, no complete gap can open in $\bar{\rho}(\varepsilon)$;  
however, a possibility that a zerogap opens in $\bar{\rho}(\varepsilon)$ cannot be excluded.

If $S(\varepsilon+i0,E)$ is allowed to be discontinuous as a function of $E$, numerous types of anomaly are possible because $S(\varepsilon+i0,E)$ as a function of $E$ does not need to be analytical in the upper or lower half complex plane of $E$;
for example,  
\begin{subequations}\label{EqDeltaContinousS1} 
\begin{align}
S(\varepsilon+i0, E) =
\frac{\lambda(E)}{\varepsilon+{i}0} + \delta S(\varepsilon+i0, E),
\end{align}
where $\lambda(E)\ge 0$ and, in particular, $\lambda(0)> 0$ and 
\begin{align}\label{EqDeltaContinousS12}
\delta S(\varepsilon+i0, E) =
\left\{\begin{array}{cc}
\displaystyle
\frac{E}{|E|}\left[\sqrt{E^2 + (\epsilon_{\rm g}/2)^2} - |E|\right], & E\ne 0 \vspace{2pt},\\
0, & E=0 ,
\end{array} \right. 
\end{align}
\end{subequations}
\begin{align}\label{EqDeltaContinousS2}
S(\varepsilon+i0, E) =
\frac{(\epsilon_{\rm g}/2)^2}{\varepsilon+{i}0 + E},
\end{align}
and so on.
If $S(\varepsilon+i0, E)$ is given by Eq.\hskip2pt(\ref{EqDeltaContinousS1}) or (\ref{EqDeltaContinousS2}), a complete gap as large as $\epsilon_{\rm g}$ opens, 
even if the bandwidth is infinite. 
A possibility that a complete gap opens in $\bar{\rho}(\varepsilon)$ cannot be excluded.
A necessary condition for a complete gap to open is that
$S(\varepsilon+i0, E=0)$ or $S(\varepsilon+i0, k=\pm k_{\rm F})$ has a pole at $\varepsilon=0$ and $S\left[\varepsilon+i0, E(k)\right]$ is discontinuous as a function of $E$ or $k$.

The single-band model examined above can be substituted for a two-band model.
We define two types of creation and annihilation operators by
\begin{align}
\alpha_{k\sigma}^\dag = d_{(k+k_{\rm F})\sigma}^\dag, \quad
\alpha_{k\sigma} = d_{(k+k_{\rm F})\sigma}, \quad 
\beta_{k\sigma}^\dag = d_{(k-k_{\rm F})\sigma}^\dag, \quad
\beta_{k\sigma} = d_{(k-k_{\rm F})\sigma},
\end{align}
for $- k_{\rm F}< k < k_{\rm F}$; $\alpha$ and $\beta$ electrons correspond to right-going and left-going ones, respectively, in the single-band model.
The one-body term is substituted for
\begin{align}
{\cal H}_0 = \sum_{k\sigma} \hbar v_{\rm F} k\left(
\alpha_{k\sigma}^\dag \alpha_{k\sigma} - \beta_{k\sigma}^\dag \beta_{k\sigma}
\right).
\end{align}
The two-body term is decomposed into several terms according to the numbers of $\alpha_{k\sigma}^\dag$, $\alpha_{k\sigma}$, $b_{k\sigma}^\dag$, and $b_{k\sigma}$, some of which are the so called umklapp terms.
This two-band model is none other than the Tomonaga-Luttinger model with umklapp terms.\cite{Tomonaga,Luttinger,Mattis}
Since the two-band model is a continuous model the same as the original single-band model, the Kondo-lattice theory is inapplicable to it.
Since each of the numbers of $\alpha$ and $\beta$ electrons is not a  conserved quantity because of the two-body term,
the Green function is given, in general, by
\begin{align}
{\cal G}_\sigma(i\varepsilon_l,k) &=
\left(\hskip-3pt\begin{array}{cc}
i\varepsilon_l - \hbar v_{\rm F} k - S_{\alpha\alpha\sigma}(i\varepsilon_l,k) & 
- S_{\alpha\beta\sigma}(i\varepsilon_l,k)\\
- S_{\beta\alpha\sigma}(i\varepsilon_l,k) & 
\hskip-5pt
i\varepsilon_l + \hbar v_{\rm F} k - S_{\beta\beta\sigma}(i\varepsilon_l,k)
\end{array}\hskip-3pt\right)^{-1} . 
\end{align}
If the correspondence to the original single-band model is seriously considered, the interband terms have to vanish such that
\begin{subequations}\label{EqEqualSS}
\begin{align}
S_{\alpha\beta\sigma}(i\varepsilon_l,k) =
S_{\beta\alpha\sigma}(i\varepsilon_l,k) =0, 
\end{align}
and the intraband terms have to satisfy
\begin{align}
S_{\alpha\alpha\sigma}(i\varepsilon_l,k) 
=S_{\beta\beta\sigma}(i\varepsilon_l,-k) 
=\Sigma_{\sigma}\bigl[i\varepsilon_l,E(k+k_{\rm F})\bigr]-\mu. 
\end{align}
\end{subequations}
The density of states of this two-band model is equal to that of the original single-band model, provided that Eq.\hskip2pt(\ref{EqEqualSS}) is satisfied.
Thus, a necessary condition for a complete gap to open is that $S_{\alpha\alpha\sigma}(\varepsilon+i0,k)$ has a pole at $\varepsilon=0$ at least for $k=0$ and $S_{\alpha\alpha\sigma}(\varepsilon+i0,k)$  is discontinuous as a function of $k$.

It is desirable to elucidate how discontinuous $S_{\alpha\alpha\sigma}(\varepsilon+i0,k)$ is as a function of $k$ in the gapped phase of the Tomonaga-Luttinger model, which corresponds to the gapped phase of the boson model.\cite{Solym}
Since it is surprising that the self-energy as a function of wave number $k$ is discontinuous, it is desirable to elucidate what is actually responsible for the discontinuity.
It is also desirable to elucidate whether the residual entropy is zero or nonzero in the gapped phases;
the divergence of ${\rm Im}S_{\alpha\alpha\sigma}(+i0, 0)$  implies or at least suggests that there are degenerate zero-energy bosonic excitations, which 
are responsible for the divergent ${\rm Im}S_{\alpha\alpha\sigma}(+i0, 0)$.

\end{document}